\documentclass[11pt,a4paper]{article}
\pdfoutput=1
\usepackage{jheppub}
\usepackage{subfig}
\usepackage{graphicx}

\usepackage{bm} \usepackage{slashed} \usepackage{xcolor} \DeclareMathOperator{\tr}{tr}  
\title{Differential equations from unitarity cuts: nonplanar hexa-box integrals}
\author[a]{Samuel Abreu,}
\author[a]{Ben Page,}
\author[b]{Mao Zeng}
\affiliation[a]{Physikalisches Institut, Albert-Ludwigs-Universitat Freiburg,\\ Hermann-Herder-Straße 3, D–79104 Freiburg, Germany}
\affiliation[b]{Mani L.\ Bhaumik Institute for Theoretical Physics,\\
Department of Physics and Astronomy, University of California,\\
475 Portola Plaza, Los Angeles, CA 90095, USA}
\emailAdd{abreu.samuel@physik.uni-freiburg.de}
\emailAdd{ben.page@physik.uni-freiburg.de}
\emailAdd{zengmao@physics.ucla.edu}
\abstract{We compute $\epsilon$-factorized differential
equations for all dimensionally-regularized integrals of the
nonplanar hexa-box topology, which contribute for instance to
2-loop 5-point QCD amplitudes. A full
set of pure integrals is presented.
For 5-point planar topologies,
Gram determinants which 
vanish in $4$ dimensions are used to build
compact expressions for
pure integrals.
Using unitarity 
cuts and computational algebraic geometry,
we obtain a compact IBP system
which can be solved
in 8 hours on a single CPU core, overcoming a major
bottleneck for deriving the differential equations.
Alternatively, assuming prior knowledge of the
alphabet of the nonplanar hexa-box, we reconstruct analytic
differential
equations from 30 numerical phase-space points, making 
the computation almost trivial with current techniques. We solve the differential equations to obtain the values of the master integrals at the symbol level.
Full results for the differential equations and solutions are included as supplementary material.}
\preprint{FR-PHENO-2018-009, UCLA/18/TEP/105} 
\arxivnumber{1807.11522}
\begin{document}
\maketitle

\section{Introduction}
\label{sec:intro}
\makeatletter{}

In the coming years, Run 2 of the LHC and the future 
high-luminosity LHC will accumulate large datasets and
deliver high-precision experimental measurements.
These must be
matched by high-precision theoretical predictions for the 
relevant cross sections, at NNLO for
a number of observables. A fundamental ingredient in
obtaining these predictions is the evaluation of two-loop 
Feynman integrals with several external legs. 
Feynman integrals are also crucial for studying
scattering amplitudes in supersymmetric theories, where there 
has been great progress in constructing loop integrands but
much less
is known about integration, especially beyond the planar limit. For instance, while the nonplanar
four-particle amplitude is known at three-loop order 
\cite{Henn:2016jdu}, at five-points only the integrand is known
at two-loops \cite{Carrasco:2011mn,Bern:2015ple}.

A problem that has recently attracted much attention is the
calculation of massless two-loop five-point amplitudes.
The two-loop five-point amplitudes for pure
Yang-Mills theory have been recently computed in the
planar limit, 
first for the special case of all-plus external helicities 
\cite{Badger:2013gxa,Gehrmann:2015bfy, Dunbar:2016aux} and 
subsequently for arbitrary helicities 
\cite{Badger:2017jhb, Abreu:2017hqn}.
The calculation of loop amplitudes requires both the evaluation
of the integrand, which for the five-point amplitudes mentioned
above has often been performed in the framework of generalized
unitarity \cite{Bern:1994zx,Bern:1994cg,
Bern:1997sc, Britto:2004nc}, and the evaluation of Feynman
integrals. An important property of dimensional regularization 
is that total derivatives with respect to the loop momenta
of Feynman integrands in momentum 
space integrate to zero (boundary terms always vanish).
This gives rise to powerful integration-by-parts (IBP) 
identities \cite{Chetyrkin:1981qh, Laporta:2001dd,
Laporta:1996mq}, which relate integrals with a specific set of
propagators but with arbitrary numerators and different integer
exponents of propagators to a basis of so-called 
\emph{master integrals}. 
Using these relations, an amplitude can be reduced to 
a linear combination of master integrals that must then be
evaluated. All  
master integrals required for planar
massless five-point two-loop
amplitudes have been evaluated 
analytically \cite{Papadopoulos:2015jft,Gehrmann:2018yef}. 
Beyond the planar limit, the amplitudes are as yet unknown and
require in particular the evaluation of nonplanar
five-point two-loop integrals. Some
master integrals have been recently
evaluated using bootstrap and (super-) conformal symmetry 
methods \cite{Chicherin:2017dob, Chicherin:2018ubl}, with additional hidden symmetry properties explored in \cite{Bern:2018oao, Chicherin:2018wes}, but full
results for the nonplanar hexa-box and double-pentagon 
topologies are still missing.

A very successful method for evaluating 
dimensionally-regularized master integrals is through the
solution of differential equations with respect to kinematic
invariants \cite{Kotikov:1990kg, Bern:1993kr, Remiddi:1997ny, 
Gehrmann:1999as, Argeri:2007up,Henn:2013pwa}. IBP relations also
play a central role in this method. Indeed, the differential
equations are obtained by differentiating Feynman 
integrals, and rewriting the resulting expression in terms of
the basis of master integrals one is trying to compute. This
last step is done with the help of IBP relations. The set of IBP
relations that is needed for differential equations is in 
general simpler to obtain when compared to the one required for
the evaluation of an amplitude
because the integrals that must 
be reduced are simpler 
(indeed, only recently have IBP-reduction tables for planar
five-point massless amplitudes been obtained 
\cite{Boels:2018nrr,Chawdhry:2018awn}). 
Despite this, for nonplanar five-point
two-loop integrals the IBP-reduction step is still the
bottleneck in computing the full set of master integrals from 
the differential-equation approach.

In this paper, we present a method for constructing the
differential equations which overcomes this issue.
The main result is an approach to obtain a more compact IBP
system than the one in used in more conventional methods
\cite{Anastasiou:2004vj, vonManteuffel:2012np, Lee:2012cn, 
Smirnov:2014hma, Maierhoefer:2017hyi}
which is thus easier to solve.
It builds on
the recent observation that, in conjunction with computational
algebraic 
geometry tools, unitarity-based methods
can be applied to construct IBP relations 
\cite{Gluza:2010ws, Ita:2015tya, Larsen:2015ped,
 Georgoudis:2016wff, Abreu:2017xsl, Abreu:2017hqn, 
 Boehm:2017wjc, Boehm:2018fpv} and differential equations 
\cite{Frellesvig:2017aai, Zeng:2017ipr, Bosma:2017hrk}.
Our approach is based on the construction of \emph{unitarity-compatible IBP relations} 
\cite{Gluza:2010ws, Ita:2015tya, Larsen:2015ped,
Georgoudis:2016wff, Abreu:2017xsl, Abreu:2017hqn, 
Boehm:2017wjc, Boehm:2018fpv}, which allow us to control the 
powers of the propagators in the integrals appearing when
generating the IBP relations. It is made even more
efficient by working on specific unitarity cuts and then
combining the results obtained on each of the cuts, in a
procedure similar to the one used in 
\cite{Larsen:2015ped, Boehm:2018fpv}. As an example of our
approach, we compute the differential equations for 
the nonplanar hexa-box topology. 
The differential equations for the
maximal-cut integrals of this topology were previously
derived by one of the authors using unitarity-based 
methods~\cite{Zeng:2017ipr}. Here we present results for the 
differential equations of the uncut integrals.

Although any basis of master integrals is in principle
equivalent, it is well known that a basis of so-called 
\emph{pure master integrals} is more convenient for solving the
differential equations, as the dependence on the dimensional
regularization parameter $\epsilon=(4-d)/2$ completely
factorizes \cite{Henn:2013pwa}. 
We thus construct such a basis and use it to write the
differential equations. 
Explicit expressions were known in the literature for most of 
the pure master integrals we required,
except for some 5-point examples.\footnote{
During the final stages of the preparation of this paper, a
basis of pure integrals for the planar sectors was
presented in \cite{Gehrmann:2018yef}. We present an alternative
construction.
}
Pure integrals are often constructed from the analysis of the
integrand in strictly 4 dimensions. In this paper we observe
that numerators constructed from specific Gram
determinants that vanish in 4 dimensions can also be used. 
They lead to very
compact expressions for the numerators, and we give explicit
expressions for all planar five-point topologies, even beyond
the ones required for the present calculation.
The differential
equations in our pure basis are given in the ancillary file 
{\tt diffEqMatrices.m}.

We also propose an alternative and much more direct way of
obtaining the differential equations of the nonplanar hexa-box, 
which relies on the knowledge of its so-called 
\emph{alphabet}~\cite{Goncharov:2010jf}
and of a pure basis. In this approach, the analytic forms of the
differential equations are highly constrained, leaving a small 
number of free parameters that are fixed by numerical fitting,
with IBP reduction computed at numerical values of the 
kinematic invariants. This makes the computational resources
required trivial.

To validate our results, we compute the solution of the
differential equations at the symbol~\cite{Goncharov:2010jf}
level by constructing a generic solution, imposing the 
so-called \emph{first-entry condition}~\cite{Gaiotto:2011dt} on
the symbol, and finally computing a single trivial integral that
allows to determine all initial
conditions required at the symbol level.

The paper is structured as follows. In Section \ref{sec:ibp}, we
describe our approach to generating the
unitarity-compatible IBP relations required
for computing differential equations, and the application to the
case of the nonplanar hexa-box.
In Section \ref{sec:de}, we discuss the use of the IBP reduction
results to obtain the
differential equations in dimensional regularization, the
transformation into a pure basis, how we construct the pure
basis, and our alternative method of constructing the
differential equations.
In Section \ref{sec:results} we discuss our results for the
differential equations of the nonplanar hexa-box and their
solution at symbol level.
In Appendix \ref{sec:sector}
we describe some further details of our implementation of IBP
reduction, namely the generation of IBP identities
sector-by-sector. In Appendix \ref{sec:momTwist} we discuss an
alternative set of variables leading to a rational alphabet.
Finally, in Appendix  \ref{sec:basis} we tabulate all the $73$
pure integrals in our pure basis.

\section{Unitarity-compatible integration-by-parts reduction}
\label{sec:ibp}
\makeatletter{}
A crucial step in constructing differential equations for a
Feynman integral is obtaining the associated set of
integration-by-parts (IBP)
relations. These are required to rewrite the derivatives of the
integrals in terms of a set of master integrals. This crucial 
step is currently the main bottleneck in calculating new Feynman
integrals
through the method of differential equations.

\begin{figure}
  \centering
  \includegraphics[width=0.5\textwidth]{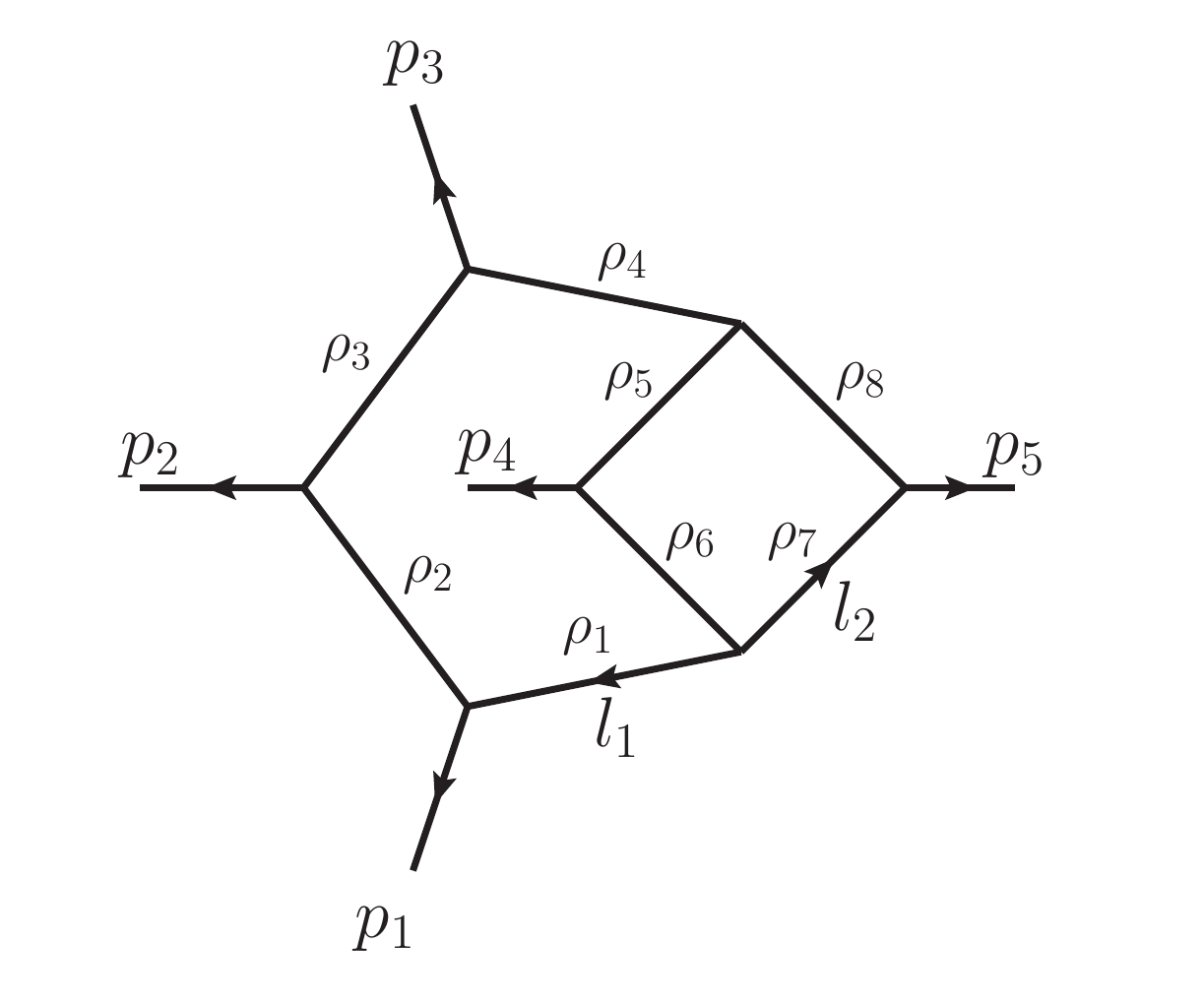}
  \caption{The nonplanar hexa-box topology. We consider the case
  with all external and internal lines massless.}
 \label{fig:hexabox}
\end{figure}

In this paper we describe an approach to tackle this issue.
The improvement we present is based on a more efficient
construction of the IBP relations. We focus on the
application of the approach to the construction of a system of
differential equations for the nonplanar
hexa-box topology shown in fig.~\ref{fig:hexabox}, which has so
far not been achieved through more standard techniques. 
This
serves as an illustration of the potential of our approach, 
which is completely generic and applicable beyond this example.

\subsection{Loop momentum parametrization}
\label{subsec:param}

Let us begin by defining our conventions to describe the nonplanar
hexa-box topology.
We choose the five independent kinematic invariants to be
\begin{equation}
\label{eq:defSij}
s_{12}, s_{23}, s_{34}, s_{45}, s_{51},
\end{equation}
defined as
\begin{equation}
s_{ij} = (p_i + p_j)^2 \, . 
\end{equation}
We always set
\begin{equation}
s_{45} = 1
\end{equation}
in our calculations to eliminate one kinematic scale, which 
amounts to normalizing all variables to an overall scale.
The analytic dependence on this overall scale can always be
recovered from dimensional analysis.

With the conventions of fig.~\ref{fig:hexabox},
the inverse propagators are written as
\begin{align}
\label{eq:hexaboxProps}
&\rho_1 = l_1^2, \quad \rho_2 = (l_1 - p_1)^2, \quad \rho_3 = (l_1 - p_1 - p_2)^2, \quad \rho_4 = (l_1 + p_4 + p_5)^2, \nonumber \\
&\rho_5 = (l_1 + l_2 + p_4)^2, \quad \rho_ 6 = (l_1 + l_2)^2, \quad \rho_7 = l_2^2, \quad \rho_8 = (l_2-p_5)^2 \, .
\end{align}
There are three `irreducible numerators' that cannot be written
as a linear combination of inverse propagators. 
We choose to define them as
\begin{equation}
\rho_9 = (l_2 + p_1)^2, \quad \rho_{10} = (l_2 + p_1 + p_2)^2, \quad \rho_{11} = (l_2 + p_1 + p_2 + p_3)^2 \, .
\end{equation}

Our approach to the construction of the IBPs
is inspired by generalized unitarity and thus heavily relies on
putting some of the propagators `on-shell'. In the context of
this paper, putting a propagator $1/\rho$
on-shell simply means we set
its inverse $\rho$ to zero when it appears in the numerator.
It is thus convenient to parametrize the loop momenta in terms 
of the inverse propagators and irreducible numerators
\cite{Cutkosky:1960sp, Baikov:1996rk, Baikov:1996iu, 
Grozin:2011mt, Ita:2015tya, Larsen:2015ped}
\begin{equation}
\label{eq:varsNat}
\rho_1, \,\rho_2, \,\, \dots \,\, ,\, \rho_{11} \, .
\end{equation}
We shall call these \emph{natural variables} 
(in recent literature, they are also known as Baikov variables).
There is an invertible linear map between the natural
variables and the following eleven dot product, which we call 
the \emph{canonical variables},
\begin{equation}
\label{eq:varsCanon}
l_1 \cdot l_1,\quad l_2 \cdot l_2,\quad l_1 \cdot l_2\,, 
\quad l_i \cdot p_A\,, \qquad \text{where }
\quad i=1,2\,, 
\quad \text{and} \quad A=1,2,3,4 \,.
\end{equation}
Therefore, any expression that is a polynomial in the canonical
variables, such as the numerator of a loop integrand, is also a
polynomial in the natural variables. It then becomes trivial to
impose unitarity cuts on such an expression. For example, the 
maximal cut of a numerator of the nonplanar hexa-box 
topology is 
simply computed by setting
$\rho_1 = \rho_2 =\dots =\rho_8=0$, leaving a polynomial in the
three irreducible numerators.

\subsection{IBP relations with controlled propagator powers}
In dimensional regularization, integrals of total derivatives
with respect to the loop momenta of a
Feynman integral in the momentum-space representation
do not have boundary terms. This leads to the so-called IBP
relations, which relate Feynman integrals with different powers
of propagators. For instance, for the 
nonplanar hexa-box integrals, IBP relations are
obtained from 
\begin{equation}
\label{eq:defIBP}
0 = \int d^d l_1 \int d^d l_2 \, \frac{\partial}
{\partial l_i^\mu} \frac{v_i^\mu}{\prod_{j=1}^8 \rho_j},
\end{equation}
for a generic vector $v_i^\mu$,
where there is an implicit summation over the loop momentum 
label $i=1, 2$. We can expand $v_i^\mu$ as
\begin{equation}
v_i^\mu = v_{ij} \, l_j^\mu + v_{iA} \, p_A^\mu,
\end{equation}
where the implicit summation ranges are $j=1,2$ 
and $A=1,2,3,4$.
Therefore the  $v_i^\mu$ have a total of $2 \times (2+4) = 12$ 
coefficients $v_{ij}$ and $v_{iA}$. We require all $12$
coefficients to be polynomials in the loop-momentum components
such that the object we obtain after computing the derivative is
still a Feynman integral.

For generic vectors $v_i^\mu$, the action of the derivative will
naturally lead to propagators with increased power (squared
propagators in the case of eq.~\eqref{eq:defIBP}).
Therefore, conventional implementations of integration-by-parts
relations \cite{Anastasiou:2004vj, Lee:2012cn, 
vonManteuffel:2012np, Smirnov:2014hma, Maierhoefer:2017hyi}
involve `auxiliary integrals' with raised 
propagators powers. Although one is a priori not interested in
the reduction of these auxiliary integrals, in conventional
approaches one is forced to
also reduce them in order to have a complete IBP system. This
leads to a proliferation of terms and can make solving the IBP
system a very complicated task. 

A method that avoids the appearance of auxiliary integrals
was proposed in ref.~\cite{Gluza:2010ws}. It is based on the
observation that these will not be generated if one demands
that the vector $v_i^\mu$ satisfy
\begin{equation}
\label{eq:GKK}
v_i^\mu \frac{\partial}{\partial l_i^\mu}\, \rho_j = f_j \, \rho_j, \quad j=1,2, \, \dots \, , 8 \, ,
\end{equation}
where $f_j$ are arbitrary polynomials in the natural variables. 
We will call vectors
$v_i^\mu$ which solve the above equation 
\emph{IBP-generating vectors}. We stress that the above 
condition applies to the inverse propagators but not to
the irreducible numerators, i.e., in our example it applies to 
$\rho_1, \rho_2, \, \dots \, , \rho_8$ but not to
$\rho_9, \rho_{10}, \rho_{11}$. It is clear that, when
using such an IBP-generating vector in
eq.~\eqref{eq:defIBP}, no terms with squared propagators will be
generated on the right-hand side of the equation, 
and one thus obtains a
relation between integrals with unit powers of the propagators.

Note that this procedure also allows us to deal with integrals that
have squared propagators (or, in fact, a propagator raised to
an arbitrary positive power). Indeed, consider the example
\begin{equation}
\label{eq:defIBPWithDots}
0 = \int d^d l_1 \int d^d l_2 \, \frac{\partial}{\partial
l_i^\mu} 
\frac{v_i^\mu}{\rho_1 \rho_2 \dots \rho_k^2 \dots \rho_8}.
\end{equation}
If the vectors $v_i^\mu$ satisfy the condition in 
eq.~\eqref{eq:GKK}, then the right-hand side
 will involve integrals with 
the propagator $1/\rho_k$ raised to power two or less, and the
remaining propagators raised to power one or less. Also in this
case, we find that the IBP-generating vectors allow to control
the proliferation of auxiliary integrals. We note that this case
is important if one wants to use IBP relations for constructing
differential equations, as integrals with squared propagators can
be generated when differentiating an integral with respect to an
external kinematic invariant.

In summary, using IBP-generating vectors, we are 
able to generate IBP relations in which we have control over
the power of the propagators of the integrals. In particular, we
never generate integrals with higher powers of propagators.
IBP relations obtained in this approach are called 
\emph{unitarity-compatible IBP relations}. 
These IBPs lead to a more
compact IBP system that is thus easier to solve.\footnote{One
might wonder whether one can always find a complete IBP system
using only such IBP relations. Although we do not have a general
proof, we have not yet found a case where this is not 
possible.}

The final question we must address is how we construct the
IBP-generating vectors.
A full set of solutions to eq.~\eqref{eq:GKK} can be obtained 
from computational algebraic geometry~\cite{Gluza:2010ws, 
Larsen:2015ped, Zhang:2016kfo, Abreu:2017xsl, Abreu:2017hqn, 
Boehm:2018fpv}, computational linear algebra~\cite{
Schabinger:2011dz}, and in some cases, analytic formulas from 
loop-by-loop considerations~\cite{Ita:2015tya} and dual 
conformal symmetry~\cite{Bern:2017gdk}. In this paper, we adopt 
the method of ref.~\cite{Abreu:2017hqn}, which directly solves 
eq.~\eqref{eq:GKK} after rewriting the 
left-hand side in terms of natural 
variables, using algorithms implemented in the computational algebraic
geometry package {\tt SINGULAR}~\cite{DGPS}. In addition, we identify
and remove redundant IBP-generating vectors found by {\tt SINGULAR}, by reducing the vectors against each other at random numerical values of the kinematic invariants $s_{ij}$, modulo an arbitrary large prime number. We only retain vectors that are not reduced to zero by other vectors. This ad hoc procedure, similar to the one in ref.~\cite{Boehm:2018fpv} but without Groebner-basis computations, significantly reduces the number of IBP-generating vectors.

\subsection{Simplification on unitarity cuts}
\label{subsec:simp}

In this subsection, we discuss IBP
reduction on a spanning set of non-maximal cuts before full
results are obtained from merging the results computed on
specific cuts, in a procedure similar to the one used in
refs.~\cite{Larsen:2015ped, Boehm:2018fpv}.
This approach further simplifies the construction
of the IBP relations.

We first introduce a compact notation for integrals with 
generic powers of propagators and numerators
(specialized to the nonplanar hexa-box topology),
\begin{equation}
\label{eq:defFNotation}
F[\vec \nu] = 
F[ \nu_1, \nu_2, \, \dots \, , \nu_{11} ] 
= \left(\frac{e^{\gamma_E\epsilon}}{i\pi^{d/2}}
\right)^2
\int d^d
l_1 \int d^d l_2 \prod_{j=1}^{11} \rho_j^{-\nu_j} \, .
\end{equation}
Here, the $\nu_i$ denote the powers of propagators and 
numerators. We include a standard overall normalisation factor
that depends on the dimensional regulator $\epsilon$ 
(we take
$d=4-2\epsilon$) and the Euler–Mascheroni constant $\gamma_E$.
The indices $\nu_{9}, \nu_{10}, \nu_{11}$ are required to be
non-positive, since the three irreducible numerators $\rho_9,
\rho_{10}, \rho_{11}$ never appear in the denominator with
positive powers. On the other hand, the indices 
$\nu_{1}, \nu_{2}, \,\dots \, , \nu_{8}$ are allowed to be any 
integers. Indeed, we include integrals of \emph{sub-topologies} 
with a proper subset of the propagators where the missing 
inverse propagators can appear in the numerator with 
positive powers. Every IBP relation is a linear relation 
between the different $F[\vec \nu]$ integrals in 
eq.~\eqref{eq:defFNotation} with different sets of $\nu_i$ 
indices.

In the following, we will consider 
the generalized unitarity cut of a
subset $\Delta$ of the propagators of a given $F[\vec \nu]$,
\begin{equation}
\Delta \subseteq \{1,2, \dots, 8\}\,.
\end{equation}
For the purpose of
this paper, it is sufficient to note that
\begin{equation}
\label{eq:defCutOnF}
F[\vec \nu] \big|_{\Delta} = 0 \quad \text{if} \quad
\nu_j \leq 0 \ \text{for any } j \in \Delta \, .
\end{equation}
In other words, the cutting operation sets to zero any integral
that does not have all cut propagators raised to nonzero
positive powers.

As defined above, the cutting operation is a linear operation.
If one has constructed an IBP relation for uncut integrals, one
can then apply a cut to the relation and obtain the IBP
relation between the cut integrals. Since some integrals are set
to zero by the cutting operation, the cut relation will be
simpler (i.e., involving fewer integrals) than its uncut
counterpart. One thus expects that it should be simpler to
compute IBP relations on a given cut, and this is indeed true. 
In particular, it is simpler to construct IBP-generating 
vectors by solving eq.~\eqref{eq:GKK} subject to cut
conditions. One thus obtains IBP relations modulo integrals that
do not depend on the cut propagators.

To choose which cuts to consider, we construct all subsets of
propagators and check if, when removing any extra propagator,
the corresponding Feynman integral becomes scaleless. If that is
the case, we keep the cut associated with that subset. These cuts form a \emph{spanning set of cuts}, and
an explicit set for the nonplanar hexa-box will be
given in the next section. In practice, working on a cut means 
we have to construct IBP-generating vectors that depend on fewer
variables, which can greatly simplify their calculation using
computational algebraic geometry.

Having constructed an IBP relation 
on a given cut, there are two types of terms we need to
worry about not capturing. The first type corresponds to 
integrals that only have a subset of the cut propagators. Given
the way we chose the cuts, these are
scaleless, vanishing integrals, and thus they do not 
modify the IBP relation. The
second type of terms that we miss on a given cut are
terms with a subset of the cut propagators and some other
propagators. These terms will be captured by another cut in the
spanning set. By \emph{merging} the information on the spanning 
set of cuts, we
construct the full IBP-reduction table which 
reduces any uncut integral
to a set of master integrals. The coefficient of a particular
master integral is computed from any cut which does not set 
this master integral to zero.

When we use eq.~\eqref{eq:defIBPWithDots} to generate IBP
relations with doubled propagators $1/\rho_k^2$, care must be
taken when we work with a cut that includes the $k$-th
propagator. We temporarily treat the propagator $1/\rho_k$ as
uncut when constructing IBP-generating vectors from 
eq.~\eqref{eq:GKK}. After the associated IBP relations have been
generated, we re-impose the cut conditions and set any integral
to zero if the two powers of $1/\rho_k^2$ are completely 
canceled in the integral. More explicitly, we set to zero any
integral for which the $k$-th index in
eq.~\eqref{eq:defFNotation} is strictly non-positive,
rather than being $1$ or $2$.

\subsection{Implementation for the 2-loop 5-point nonplanar hexa-box}
\label{subsec:implementation}

In this section we give the details of the implementation
of the general formalism of the previous subsections to the
calculation of the IBP relations required for the
differential equations of
the nonplanar hexa-box. We need
to generate two kinds of IBP relations:
\begin{enumerate}
\item IBP relations involving only integrals with no doubled propagators.
\item IBP relations allowing for integrals with a doubled
propagator
$1/\rho_2^2$.
\end{enumerate}
Furthermore, it is sufficient for our purposes to 
generate the IBP relations at numerical values of $d$. The
reasons why this is sufficient will be made clear in the next
section.

We perform IBP reduction on 8 triple cuts and 2 quadruple cuts,\footnote{We thank Julio Parra-Martinez for 
graph theory code for identifying scaleless Feynman diagrams.}
containing the following subsets of propagator
lines as labeled by eq.~\eqref{eq:hexaboxProps},
\begin{align}
\label{eq:cutList}
& \Delta_1 = \{1,5,8\}, \quad \Delta_2 = \{2, 5, 7\}, \quad \Delta_3 = \{2,5,8\}, \quad \Delta_4 = \{2,6,8\}, \nonumber \\
& \Delta_5 = \{3,5,7\}, \quad \Delta_6 = \{3,6,7\}, \quad \Delta_7 = \{3,6,8\}, \quad \Delta_8 = \{4,6,7\}, \nonumber \\
& \Delta_9 = \{1,4,5,7\}, \quad \Delta_{10} = \{1,4,6,8\} \, .
\end{align}
When generating IBP relations with a doubled propagator
$1/\rho_2^2$ using eq.~\eqref{eq:GKK}, 
the propagator $1/\rho_2$ must be uncut, i.e.\
deleted from the cut lines. The following double cuts are then
also needed,
\begin{equation}
\tilde \Delta_2 = \{5,7\}, \qquad \tilde \Delta_3 = \{5,8\}, \qquad \tilde \Delta_4 = \{6,8\} ,
\end{equation}
and the triple cuts are re-imposed on the resulting IBP relations before we solve the IBP linear systems.

The IBP relations are in fact generated sector-by-sector.
This involves solving eq.~\eqref{eq:GKK} for every sub-topology
supported on the cut, and details of this procedure
are presented in Appendix
\ref{sec:sector}. The two kinds of IBP relations are combined 
into a single linear system which we solve on each
cut using a private code written in Wolfram Mathematica, 
calling {\tt Fermat}~\cite{lewis2008computer} through 
A.V.~Smirnov's 
{\tt FLink} (part of the {\tt FIRE5} package 
\cite{Smirnov:2014hma}).
We first set the kinematic variables $s_{ij}$ to numerical 
values, and perform Gaussian elimination with a partial (row) 
pivoting scheme that preserves the sparsity of the linear 
system. The pivotal rows correspond to a minimal set of 
independent IBP relations. We then perform Gaussian elimination 
with identical pivot choices but with full analytic dependence 
on the $s_{ij}$, starting from the independent IBP relations 
identified during the previous numerical step.
In agreement with ref.~\cite{Boehm:2018fpv}, we find that there
are 73 master integrals (75 after IBP reduction, and 73 after
accounting for symmetries).

The linear systems obtained for the quadruple cuts $\Delta_9$
and $\Delta_{10}$ in eq.~\eqref{eq:cutList} are relatively 
small and require only a few minutes each to be solved. For 
the linear systems of the remaining triple cuts, the 
computation times range from about half an hour to nearly two 
hours. The total time used to solve these linear systems is 
about 8 hours, using one CPU core.\footnote{The CPU is an Intel 
Xeon E7-8890 with a clock speed of 2.2 GHz and cache size of 60
MB. Our
program only uses one core out of the 96 available on the node.}
In the end, the results for the $10$ cuts are merged to 
produce the full IBP reduction table for all integrals with up 
to rank-3 numerators and up to one doubled propagator 
$1/\rho_2^2$. By
applying a graph symmetry we obtain from it the reduction table
for integrals with the doubled propagator $1/\rho_3^2$.

\section{Differential equations and basis of pure integrals}
\label{sec:de}
\makeatletter{}
\subsection{Kinematic derivatives and momentum routing}
\label{sec:kinDeriv}

As discussed above, it is sufficient to fix $s_{45} = 1$ and
compute the partial derivatives of the master integrals with respect to 
the other kinematic variables, $s_{12},s_{23}, s_{34}, s_{51}$.
These may be represented as partial derivatives with respect to the four
independent external momenta, $p_1, p_2, p_3, p_4$,
\begin{equation}
\label{eq:deriSAndDeriP} 
\frac{\partial}{\partial s_{ij}} = \beta_{ij}^{k \, \mu} \frac{\partial}{\partial p_k^\mu},
\end{equation}
where $k$ runs over $1,2,3,4$ in the implicit summation.
The calculation is made simpler by a judicious choice of 
momentum routing. 
The only invariant depending on both $p_4$ and $p_5$ is
$s_{45}$, which we have fixed to be $1$.
Therefore, as Lorentz 
transformations leave all kinematic invariants unchanged, 
infinitesimal variations in the other $4$ kinematic 
variables can
always be represented by infinitesimal variations in $p_1, p_2, 
p_3$ while keeping $p_4$ and $p_5$ unchanged. The 
combination $(p_1 + p_2 + p_3)=-(p_4+p_5)$ is also unchanged
under these transformations. We thus find that 
eq.~\eqref{eq:deriSAndDeriP} can be reduced to
\begin{equation}
\label{eq:deriSAndDeriP-v2}
\frac{\partial}{\partial s_{ij}} = \beta_{ij}^{1\, \mu} \frac{\partial}{\partial p_1^\mu} + \beta_{ij}^{2\, \mu} \frac{\partial}{\partial p_2^\mu} - \left( \beta_{ij}^{1\, \mu} + \beta_{ij}^{2\, \mu} \right) \frac{\partial}{\partial p_3^\mu} \, .
\end{equation}
It is straightforward to find the $\beta_{ij}^{1\, \mu}$ and
$\beta_{ij}^{2\, \mu}$ which give the correct kinematic
derivative while preserving the on-shellness of $p_1$, $p_2$, 
and $p_3$.\footnote{For every $s_{ij}$, there is a 
one-parameter family of solutions, due to a Lorentz 
transformation generator which leaves $p_4$ and $p_5$ 
individually invariant. We can choose an arbitrary solution out of the 
one-parameter family.}

By inspection of eq.~\eqref{eq:hexaboxProps}, we can see that 
eq.~\eqref{eq:deriSAndDeriP-v2} annihilates all inverse 
propagators except $\rho_2$ and $\rho_3$. Therefore, kinematic 
derivatives of any integral (without doubled propagators) can 
only have a doubled propagator $1/\rho_2^2$ or $1/\rho_3^2$, 
while the other six propagators will always remain with single 
powers. As discussed in the previous section, it is thus
sufficient
for us to perform IBP reduction for integrals with a doubled 
propagator $1/\rho_2^2$. From those, we can then obtain IBPs for
integrals with the doubled
propagator $1/\rho_3^2$ by applying a
graph symmetry corresponding to 
flipping fig.~\ref{fig:hexabox} upside down.

The outcome of the procedure are differential equations in the
form
\begin{equation}
\label{eq:deCompForm}
\frac{\partial I_a}{\partial s_{ij}} = (M_{ij})_{ab} \, I_b,
\end{equation}
where each $M_{ij}$ is a $73 \times 73$ matrix and the $I_a$
denote the 73 master integrals. Using
differential forms, eq.~\eqref{eq:deCompForm} can be written in 
a more compact form,
\begin{equation}
\label{eq:deFormNotation}
d \bm I = \bm M  \bm I \, .
\end{equation}
In the above equation, $\bm I$ is the column vector of the $73$ master integrals, $d \bm I$ is its exterior derivative in the kinematic variables,
\begin{equation}
d \bm I = \sum_{(ij)} \frac{\partial \bm I} {\partial s_{ij}} \, d s_{ij},
\end{equation}
and $\bm M$ is a matrix of one-forms
\begin{equation}
\label{eq:matOneForm}
\bm M_{ab} = \sum_{(ij)} (M_{ij})_{ab} \, ds_{ij} \, .
\end{equation}
A useful consistency check is the integrability condition,
\begin{equation}
d \bm M_{ac} + \bm M_{ab} \wedge \bm M_{bc} = 0,
\label{eq:integrability}
\end{equation}
which follows from applying the exterior derivative to both 
sides of eq.~\eqref{eq:deFormNotation}, or more intuitively, 
from the fact that partial derivatives commute with each other 
when we differentiate the master integrals repeatedly.

\subsection{Transformation to a pure basis}
\label{sec:pureBasisTransform}

In this section we consider a transformation into a new basis
where the differential equations take a particularly simple 
form, the so-called canonical form. 
Here we review this
concept and introduce some notation.

Under a general change of basis of master integrals,
\begin{equation}
\tilde{I}_a = T_{ab} \, I_b,
\end{equation}
or equivalently
\begin{equation}
\label{eq:tMat}
\tilde {\bm I} = \bm T \bm I,
\end{equation}
eq.~\eqref{eq:deFormNotation} becomes
\begin{equation}
d \tilde {\bm I} = \tilde {\bm M} \tilde {\bm I} = \left( \bm T \bm M  \bm T^{-1} + d\bm T \, \bm T^{-1} \right) \tilde {\bm I} \, .
\label{eq:deInNewBasis}
\end{equation}
In component notation, the new matrix is
\begin{equation}
(\tilde M_A)_{ab} = T_{ac} \,(M_A)_{cd} \,T^{-1}_{db} + \frac{\partial T_{ac}}{\partial x_A} \, T^{-1}_{cb} \, .
\end{equation}

As argued in ref.~\cite{Henn:2013pwa}, for all
integrals that evaluate to multiple 
polylogarithms \cite{Goncharov:1998,Goncharov:2001}, as is the
case for the nonplanar hexa-box, we can choose a basis
of `pure' master integrals. These master integrals have the
property that the coefficients in their Laurent expansion in
$\epsilon$ have a uniform (transcendental) 
weight with no rational dependence on
the kinematic variables (we refer the reader to 
e.g.~refs.~\cite{Henn:2014qga,Duhr:2014woa} for introductory
discussions on these topics).
The weight of each coefficient is
related to the power in the $\epsilon$ expansion, such that if 
we assign weight $-1$ to $\epsilon$, each term in the Laurent
expansion has the same weight. If one finds a basis 
$\tilde {\bm I}$ where all master integrals are of this form,
then the differential equations take a very special form,
the so-called canonical form, with
\begin{equation}
\tilde {\bm M} = \epsilon \sum_\alpha \tilde M_\alpha \, d 
\log r_\alpha,
\label{eq:epForm}
\end{equation}
where the matrices $M_\alpha$ only contain rational numbers,
and all dependence on the kinematic variables appears in the 
\emph{dlog}-forms $d\log r_\alpha$. 
In component form,
\begin{equation}
\frac{\partial \tilde I_a}{\partial x_A} = \epsilon \sum_{\alpha}
\frac{\partial \log r_\alpha} {\partial x_A}  (\tilde M_\alpha)_
{ab} \,  \tilde I_b \, .
\label{eq:deCompFinal}
\end{equation}
In such a basis it is sufficient to compute the
differential equations
for a numerical value of $\epsilon$ (or equivalently $d$) since
the dependence on this variable is trivial.

Each $r_\alpha$ is a letter of the so-called 
\emph{symbol alphabet} associated with
the integrals $\tilde I_a$ \cite{Goncharov:2010jf}.
In the case of
the nonplanar hexa-box, the $r_\alpha$ are algebraic
functions of the $s_{ij}$, involving in general the square-root
of the 5-point Gram determinant. We note that
the task of writing
the solution to the differential equations in terms
of multiple polylogarithms
is greatly simplified if the alphabet is rational. This can be
achieved for this particular differential equation using the
so-called momentum-twistor variables discussed in Appendix 
\ref{sec:momTwist}. 
In particular, in these variables it becomes trivial to write a
general solution in terms of multiple polylogarithms, leaving
only the initial condition to be determined. Using known
properties of the analytic structure of Feynman integrals,
their symbol can be easily constructed from a
differential equation in canonical form. While it is not the 
same as having a complete solution of the differential equation,
the symbol encodes non-trivial information on the analytic
structure of the solution of the differential equation.

\subsection{Constructing a pure basis}

\begin{figure}[]
\centering
    \includegraphics[width=3cm]{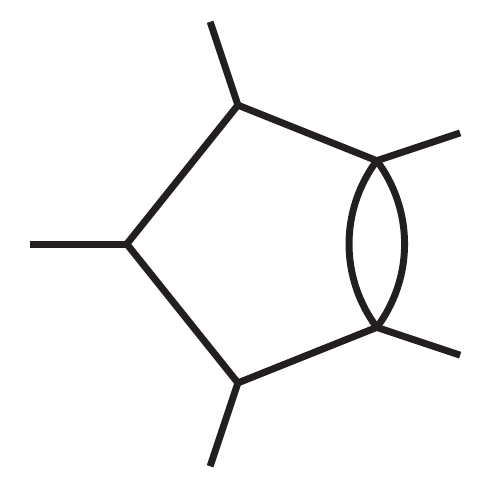}\hspace{4cm}
    \raisebox{.2cm}{\includegraphics[width=3.5cm]
    {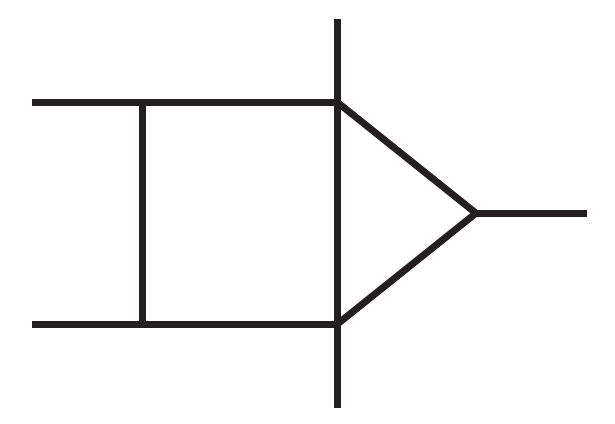}}
    \caption{Planar two-loop five-point topologies.}
    \label{fig:plan5pts}
\end{figure}

The problem of finding a pure basis of master integrals is a
complicated one, even though some strategies exist to
attempt to find such a basis in an automatic way
\cite{Lee:2014ioa,Gituliar:2017vzm,Prausa:2017ltv,
Meyer:2017joq}. 
There exist also less automatic approaches based on constructing
an integral with the appropriate analytic properties. For
instance, one might construct an integrand that can be written in a dlog
form, or an integral with 
`unit leading singularities' which is conjectured to be pure
(see \cite{Henn:2014qga} for a review of these different
approaches with examples). 

For the particular integrals that interest us in this paper, 
explicit expressions for
most pure integrals were known in the literature, in particular
in refs.~\cite{Gehrmann:2001ck,Bern:2015ple}. We had to 
construct one pure master in the nonplanar hexa-box top sector,
and all the pure master integrals for the five-point planar
integrals.\footnote{
During the final stages of the preparation of this paper, an
alternative basis of pure integrals for the planar sectors was
presented in \cite{Gehrmann:2018yef}.
}
For the latter, we observe that 
pure integrals can be obtained with numerators 
constructed from specific minors of
the Gram matrix associated with the integrand.
A basis of
pure integrals can be found in Appendix~\ref{sec:basis} and here
we give details on how the missing pure master integrals were 
constructed.

For the nonplanar sectors, we had to construct a single pure 
integral for the nonplanar hexa-box topology.
The missing pure integral, corresponding to the $\mathcal{N}_3$
numerator in table~\ref{tab:first1} in appendix~\ref{sec:basis},
was computed by constructing an integral with unit leading
singularities. We find that the numerator
\begin{equation}
\mathcal N_3 = s_{12} s_{23} \left[ \left( l_1 + p_4 \right)^2
\left( l_1 + p_5 \right)^2 - l_1^2 \left( l_1 + p_4 + p_5 \right)^2 \right] \, 
\end{equation}
on the nonplanar hexa-box has the required properties.

For the planar topologies, there are two distinct
five-point topologies, shown in
fig.~\ref{fig:plan5pts}, each with two master integrals, for which
we had to construct pure integrals.
These are relevant, for instance, for the sectors
associated with $(\mathcal{N}_{17}, \, \mathcal{N}_{18})$ and
$(\mathcal{N}_{28}, \, \mathcal{N}_{29})$ in table~\ref{tab:first1}
in Appendix~\ref{sec:basis}.
In both cases the scalar integral is pure once it is normalized
to have unit leading
singularities (for the penta-bubble
integral, there is also a simple $d$-dependent normalization).

The remaining pure master integrals are computed from 
insertions of 
loop-momentum components beyond 4 dimensions. As a motivation 
for this approach, consider the one-loop scalar pentagon   
in $6-2\epsilon$ dimensions. This integral is known to
be pure (see e.g.~refs.~\cite{Kozlov:2015kol,Abreu:2018sat} for
the
explicit differential equations in canonical form), and can be
written as a $(4-2\epsilon)$-dimensional integral with a 
nontrivial numerator insertion through simple 
dimension-shifting identities. Up to normalization, this 
numerator is
the Gram determinant of the linearly-independent momenta of the
integrand~\cite{Lee:2009dh}, which is easily related to the
square of the loop-momenta components beyond four dimensions
(below, we show this explicitly at two loops).
We apply the same principle at two loops and consider tensor
insertions of minors of the Gram matrix of the linearly
independent momenta of the integrand, denoted 
$G=G(l_1,l_2,p_1,p_2,p_3,p_4)$, and sometimes called the
Baikov matrix.
If we write the $d$-dimensional loop-momenta as
\begin{equation}
	l_i=(l_i^{[4]},l_i^{[d-4]})\,,
\end{equation}
and define
\begin{equation}
	\mu_{ij}=l_i^{[d-4]}\cdot l_j^{[d-4]}\,,
\end{equation}
then it is easy to show that, since the external
momenta live in the 4-dimensional plane,
\begin{align}\begin{split}
	\label{eq:detsAndMinors}
  \det G = 16\,(\mu_{11}\mu_{22}-\mu_{12}^2)\tr_5^2\quad
  \text{and} \quad
  \det G_{[i,j]}=16\,\mu_{ij}\tr_5^2,\qquad i,j\in\{1,2\}\,,
\end{split}\end{align}
where $G_{[i,j]}$ is the matrix obtained by deleting row $i$ and
column $j$ of $G$.
We have introduced the quantity $\tr_5$, which plays an
important role in 5-point kinematics, and can also be defined
as a trace of $\gamma$-matrices:
\begin{equation}
\label{eq:deftr5}
\tr_5 = \operatorname{Tr} \left( \gamma^5 \slashed{p_1} \, 
\slashed{p_2} \, \slashed{p_3} \, \slashed{p_4} \right) = 
\sqrt{ \det(2p_i \cdot p_j)} = i \sqrt{ \left| \det(2p_i \cdot p_j) \right|}, \quad 1\leq i,j \leq 4 \, .
\end{equation}
The above equation will be taken as the \emph{definition} of 
the branch of the square root of the 5-point Gram determinant,
as well as the chirality of the external momentum configuration
in the remaining of this paper.

By inserting specific minors from the set listed
in eq.~\eqref{eq:detsAndMinors} we are
able to construct pure integrals for the two-loop 5-point
integrals. As explicit examples, for a box-triangle and a 
penta-bubble we find that the same numerator insertion gives a suitable
choice of master integral:
\begin{equation}\label{eq:pureDouble}
\mathcal N_{17}=
\mathcal N_{28} =\frac{\tr_5}{2(4-d)}\frac{\mu_{11}}{\rho_6}\,,
\end{equation}
where $l_1$ is the loop momentum of the
box or the pentagon sub-loop respectively 
(we refer to Appendix~\ref{sec:basis} for the details on
the indexing of each master integral). We note that the factor 
of $1 / \rho_6$ means a doubled propagator 
$1 / \rho_6^2$ in the integral.

Furthermore, we observe that this approach is  
more general. Indeed, we can find pure master integrals for all planar
two-loop 5-point massless topologies using such numerators. For the double-box with five external legs, two pure
master integrals are obtained in a simple generalization of the
four-point massless case (see e.g.~\cite{Henn:2013pwa}). The
third pure master can be chosen to be
\begin{equation}
	\raisebox{-.35cm}
	{\includegraphics[width=2.5cm]{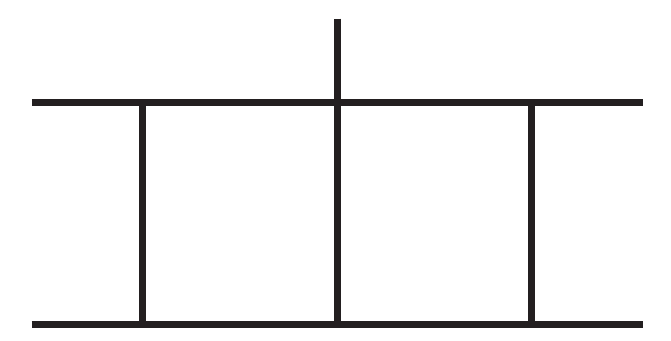}}
	\left[\mu_{12}\right]\tr_5 \,,
\end{equation}
where we wrote the term depending on the loop momenta
inside the brackets.
For the penta-box planar integral, the insertion of an
irreducible numerator (corresponding to the $t$-channel of the
sub-box) is
pure once normalized by an appropriate rational function, and
for the remaining two pure master integrals we can choose
\begin{equation}
	\raisebox{-.85cm}
	{\includegraphics[width=3cm]{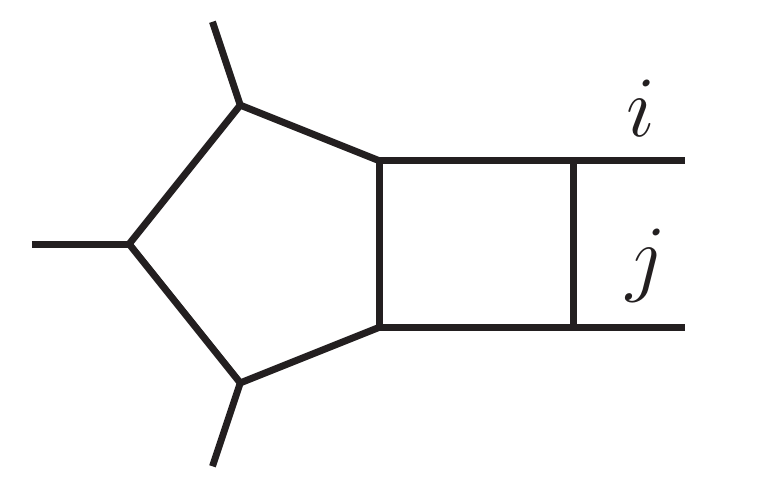}}
	\!\!\!\!\left[\mu_{12}\right]s_{ij}\tr_5\,\quad
	\text{and}\quad
	\raisebox{-.85cm}
	{\includegraphics[width=2.8cm]{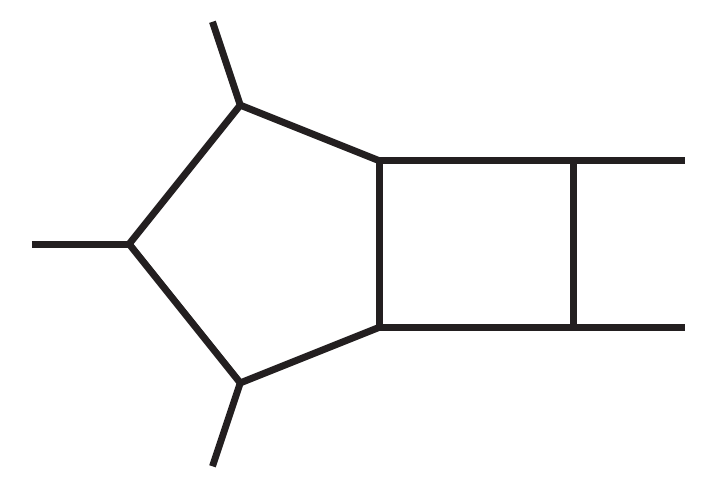}}
	\left[\mu_{11}\mu_{22}-\mu_{12}^2\right]
	\frac{d-3}{d-5}\tr_5 \,.
\end{equation}
We note that these insertions are particularly
suited to a calculation using the natural variables introduced 
in Section \ref{subsec:param}.

To finish this section, we comment on how the change of basis 
is implemented in our calculation. 
We first construct the differential equations in an arbitrary
basis of integrals, each of which is given by a single term 
$F[\vec \nu]$ in the notation of eq.~\eqref{eq:defFNotation} for
some indices $\vec \nu$. Then we construct the matrix $\bm T$
that implements the transformation to the pure 
basis, see eq.~\eqref{eq:tMat}.\footnote{
The required IBPs are constructed with the same
strategy described in Section \ref{sec:ibp}.}
Finally, the inverse matrix 
${\bm T}^{-1}$ is
computed using the software {\tt Fermat} \cite{lewis2008computer}
and Mathematica, and we obtain the differential equations in the
pure basis from eq.~\eqref{eq:deInNewBasis}.

\subsection{An alternative construction of differential
equations in canonical form}
\label{sec:alternative}

We devised an alternative
and much more direct way of
obtaining the same system of differential equations, which we now
describe. This alternative method only requires IBP reduction at
numerical  values of the $s_{ij}$, making the 
computational-resource requirements almost trivial when combined
with our
unitarity-compatible IBP reduction. However, 
it crucially relies on the a priori 
knowledge of the alphabet and is thus not generally applicable.
In the context of this paper, we have thus implemented it as a
check of the completely generic approach discussed in previous
sections.

The full alphabet for nonplanar 2-loop 5-point integrals 
has been conjectured to be composed of 31 letters given
explicitly in ref.~\cite{Chicherin:2017dob}. These are written 
in terms of the $\tr_5$ defined in 
eq.~\eqref{eq:deftr5} and the additional variables defined by
\begin{align} 
v_i \equiv s_{i, i+1}\,,\qquad
a_i \equiv v_i v_{i+1} - v_{i+1} v_{i+2} + v_{i+2} v_{i+3} - 
v_i v_{i+4} - v_{i+3} v_{i+4} ,
\end{align}
with $s_{56}$ cyclically identified with $s_{51}$.
In term of these variables, the letters are
\begin{align}\begin{split}
r_{0+i} &= v_i\,, \qquad
r_{5+i} = v_{i+2} + v_{i+3}\,, \qquad
r_{10+i} = v_{i} - v_
{i+3}, \\
r_{15+i} &= v_{i} + v_{i+1} - v_{i+3}\,,\qquad  
r_{20+i} = v_{i+2} + v_{i+3} - v_{i} - v_{i+1}\,,  \\
r_{25+i} &= \frac{a_i - \tr_5} {a_i + \tr_5}\,, \qquad
r_{31} = \tr_5 \,,
\label{eq:31symbols}
\end{split}\end{align}
 with $1 \leq i \leq 5$.
These letters are a possible choice for the
$r_\alpha$ on the right-hand side of eq.~\eqref{eq:epForm}.
Imposing $s_{45}=1$
eliminates one letter.

Using this alphabet and the pure basis of Appendix 
\ref{sec:basis}, we construct the differential
equations at 30 randomly chosen rational numerical 
phase-space
points $s_{ij}$. 
The matrices $(\tilde M_\alpha)_{ab}$
on the right-hand side of eq.~\eqref{eq:epForm}
can then be found by solving small linear
systems.\footnote{For every pair of indices 
$(a,b)$, we solve a $30 \times 30$ linear system to 
fix $(\tilde M_\alpha)_{ab}$ 
for the 
$30$ different indices~$\alpha$.} 
The use of rational numbers removes the issue of rounding errors
during the fitting procedure.
Using purely numerical data 
we thus reconstruct the
complete analytic result.

Whenever one is able to construct an ansatz
for the alphabet of a given integral and has a pure 
basis,
this method can be used to directly and efficiently construct 
the differential equations. In particular, it reduces the
reliance on the efficiency of the IBP reduction. 
We note that the test of whether a basis is pure
or not can also be done numerically, by verifying if the 
right-hand side of the differential equations 
are proportional to $\epsilon$ at a
numerical phase-space point. Furthermore, the validity of the
ansatz for the alphabet is automatically determined by whether
the linear systems have solutions.

\section{Results}
\label{sec:results}
\makeatletter{}

We have presented a completely generic method for constructing
differential equations of Feynman integrals using
unitarity-compatible IBP relations.
To show its potential, we applied it to the nonplanar
hexa-box integrals and  computed their differential equations.
This highly non-trivial example illustrates the power of our
approach as it is beyond what can currently be achieved with 
more standard approaches.
In this section we give details on how the differential 
equations were obtained and the checks that were made. In
particular, as a very strong check, we discuss the solution of
the differential equations at
the symbol level which we have compared against known results in
the literature \cite{Chicherin:2017dob}.

\subsection{Differential equations}

Following the steps described in the previous sections, 
we constructed the system of differential equations for the
nonplanar hexa-box integrals in canonical form,
using the basis of
master integrals given in Appendix~\ref{sec:basis}. 
The differential
equations are obtained at the arbitrarily chosen value of 
$d=67/13$, and working in
the finite field of cardinality $7666667$, also chosen
arbitrarily. We checked that the integrability condition 
\eqref{eq:integrability} indeed holds.
We then manipulate the expressions to obtain the explicit
decomposition in terms of the dlog-forms $d \log r_\alpha$
and the matrices $\tilde M_\alpha$ on the right-hand side of 
eqs.~\eqref{eq:epForm} and \eqref{eq:deCompFinal}.
We verified that the results remain unchanged when 
different numerical values of $d$ are used, which confirms 
the factorization of $\epsilon$ in the differential equations.
Dividing the results by $\epsilon = (4-d)/2$ and reconstructing
rational numbers from their finite-field representations, using
the extended Euclidean
algorithm, we obtain fully analytic results for the differential
equations.
We note
that the rational reconstruction from a single finite field is
only possible once the system of differential equations is
in canonical form.\footnote{The extended Euclidean algorithm
generally reconstructs the correct rational number if the
absolute values of the numerator and denominator are both 
smaller than the square root of the cardinality of the finite
field. The rational numbers in the matrices 
$\tilde M_\alpha$ are very simple,
with numerators and denominators with at most 2 digits.}
Indeed, 
if one were to attempt to reconstruct the
rational numbers in a non-pure basis, one would have to perform 
the evaluation in multiple finite fields and apply the Chinese
remainder theorem.

For the specific 
nonplanar hexa-box in figure~\ref{fig:hexabox}, we observe that
the differential equations
actually depend on
only 24 letters, instead of the 31 listed in 
eq.~\eqref{eq:31symbols}. The
remaining 7 letters will contribute once
permutations of the external legs are considered, as
required e.g.~when computing a physical scattering amplitude.
For the particular permutation of external legs
in figure \ref{fig:hexabox} we find that the following letters
are absent:
\begin{equation}
r_8, \quad r_9, \quad r_{10}, \quad r_{21}, \quad r_{22}, \quad
r_{23}, \quad r_{24}.
\end{equation}

The differential equations are derived in terms of the 
Mandelstam invariants $s_{ij}$. In these variables we observe
that, as
expected, the
$r_\alpha$ of eq.~\eqref{eq:epForm} contain square roots (of
the 5-point Gram determinant or, equivalently, $\tr_5$
as defined in eq.~\eqref{eq:deftr5}).
As already discussed, it is known that 
if we transform the results into a new set of so-called
momentum-twistor variables~\cite{Hodges:2009hk}, under which all
spinor components of external momenta have rational
components, there should be no square roots. In other words, written
in terms of the twistor variables, the $r_\alpha$ should be rational
functions. We give details of a possible choice of twistor
variables in Appendix \ref{sec:momTwist}. As a consistency check
of our calculation, we have changed to this new set of variables
and observe that the alphabet is indeed rational.  
As the most stringent check of our result, we have verified that
we reproduce exactly the same differential equations if we use
the alternative method described in Section 
\ref{sec:alternative}. As mentioned there, this is a much more
direct and computationally efficient method of obtaining the
same result, but it
relies on previous knowledge of the alphabet and of a basis
of pure master integrals. The timings required to obtaining the
differential equations through this approach are negligible, 
even compared to our approach based on unitarity-compatible IBP
relations.

The differential equations are given in the
supplementary material {\tt diffEqMatrices.m}, which contains explicit
expressions for the matrices $(\tilde M_\alpha)$ on the 
right-hand side of 
eq.~\eqref{eq:deCompFinal}, corresponding to the  
alphabet of eq.~\eqref{eq:31symbols} and the basis of Appendix
\ref{sec:basis}. The matrices are presented as two-dimensional
arrays {\tt mat[n]} in the Wolfram Mathematica 
format, with ${\tt n}=1,\ldots,31$
being the index that identifies the associated
letter.\footnote{Though the letter $v_4 = s_{45}$
disappears because we have set $s_{45} = 1$ in the calculation, 
the matrix $(M_4)_{ab}$ related to this letter can be 
fixed by the fact that the scaling dimension of every two-loop pure integral is equal to $(-4 \epsilon)$, when we choose $\mu^2=1$ as the dimensional regularization scale.} If a
 letter is absent, the corresponding matrix is the zero
matrix. As an example, the matrix $M_{31}$, corresponding to 
the letter $r_{31} = \tr_5$,
is extremely sparse, with only the following
15 nonzero entries among the $73 \times 73$ components:
\begin{align}\begin{split}
& (M_{31})_{1,1} = 2, \quad (M_{31})_{1,16} = 2, \quad (M_{31})_{2,2} = 2, \quad (M_{31})_{2,16} = -2,  \\
&  (M_{31})_{5,5} = 2, \quad (M_{31})_{5,16} = -4, \quad (M_{31})_{12,12} = 2, \quad (M_{31})_{12,16} = -4,  \\
& (M_{31})_{16,16} = -4, \quad (M_{31})_{17,17} = 2, \quad (M_{31})_{19,19} = 2,\quad (M_{31})_{24,24} = 2,  \\
& (M_{31})_{26,26} = 2, \quad (M_{31})_{28, 28} = 2, \quad (M_{31})_{30,30} = 2 \, .
\end{split}\end{align}
The matrices related to the other letters, given in the
supplementary material, also contain very simple rational numbers, whose
numerators and denominators are never larger than 16. 
The amazing simplicity of the results is the reason why we are
able to reconstruct these rational numbers from their image in a
single finite field with a 7-digit prime modulus.

\subsection{Symbols of the master integrals}

As stated in Section \ref{sec:pureBasisTransform}, once one
has obtained the differential equation in canonical form for a
Feynman integral one can easily construct its solution order by
order in $\epsilon$ at symbol level \cite{Goncharov:2010jf}. 
Indeed, constructing a generic solution at symbol level from the
differential equation is a simple exercise in combinatorics:
using the different letters one constructs a set of tensors
that are consistent with the differential equation
(see e.g.~refs.~\cite{Henn:2014qga,Duhr:2014woa} for a
pedagogical introduction). The weight-zero contributions must a
priori be determined from the initial condition, as they are
just rational numbers that vanish upon differentiation. 
However, given the iterative structure of the solution of the
differential equation, where the solution at order
$\epsilon^n$ depends on the solution at order $\epsilon^{n-1}$,
see eq.~\eqref{eq:deCompFinal}, the weight-zero constants
appear as rational coefficients of the higher-weight terms.
We can then use arguments on the analytic structure of the 
answer, such as the position of the branch points, to find
relations between the different weight-zero constants
\cite{Henn:2013pwa}. More precisely, we impose
the so-called first-entry condition~\cite{Gaiotto:2011dt} which
requires all symbol tensors to have physical branch points.
In our example, the weight-zero constants appear at order 
$1 / \epsilon^4$, and we should determine 73 such constants.
The first entry condition means that the first entries of the
tensors must correspond to the letters  $r_{0+i}$ or $r_{15+i}$ 
in eq.~\eqref{eq:31symbols}, which completely fixes all 
the 73 weight-zero coefficients up to an undetermined overall factor. 
This factor can be fixed from the
simplest master integral with three propagators, which is 
trivial to compute to all orders in $\epsilon$.\footnote{
In our normalisation, the sunrise with the numerator as in 
eq.~\eqref{eq:sunriseNorm} is $1/(4\epsilon^{4})+\mathcal{O}
(\epsilon^{-3})$.}

Following these steps, we computed the symbol of the 73 master 
integrals from our differential equations in canonical form.
As a sample result, the first integral in our pure 
basis, corresponding to the $\mathcal N_1$ numerator in 
eq.~\eqref{eq:pureIntHexabox}, has no $1/\epsilon^4$ or 
$1/\epsilon^3$ poles. The coefficient of the 
$1/\epsilon^2$ pole is given by weight-two symbol tensors built
from the letters in eq.~\eqref{eq:31symbols},
\begin{equation}
[r_{1}, r_{27}] + [r_{1}, r_{28}] + [r_{2}, r_{27}] + 
[r_{2}, r_{28}] - [r_{3}, r_{27}] - [r_{5}, r_{28}] - 
[r_{18}, r_{28}] - [r_{19}, r_{27}] \, ,
\end{equation}
which satisfies the first-entry condition stated above.

To check the validity of our results, we re-computed the 
symbols of the two nonplanar five-point integrals studied in 
ref.~\cite{Chicherin:2017dob} and found complete agreement with
their results. The symbols of all the master integrals, 
including expansion terms from order $1/\epsilon^4$ to order 
$\epsilon^0$, are given in the supplementary material {\tt symbols.m}. 
Another supplementary material {\tt masters.m} gives the pure master 
integral basis, as tabulated in Appendix \ref{sec:basis}.

\section{Conclusions}
\label{sec:conc}
\makeatletter{}
In this paper we showed that unitarity-compatible IBP relations
can be used to construct differential equations for Feynman
integrals. 
These relations reduce the size of the IBP system by avoiding 
to raise propagator powers, which leads to the proliferation of
auxiliary integrals in more standard techniques.
Although the idea of using this type of IBP relation
was introduced already some years ago \cite{Gluza:2010ws},
only recently have they been used as a powerful tool in
fully-fledged practical calculations.
In our approach to construct the differential equations,
the efficiency gained by using such compact IBP relations was
complemented by ideas from two-loop numerical unitarity 
\cite{Abreu:2017xsl, Abreu:2017hqn}: we construct analytic IBP 
relations sector-by-sector, so that we do not have to obtain 
all IBP-generating vectors simultaneously. Using our refined 
implementation, we were able to solve the IBP system needed to 
construct differential equations for all the nonplanar
hexa-box integrals in 8 hours on one CPU 
core. This required the reduction of tensor numerators of rank
up to three and up to one doubled propagator. We believe this
shows the potential of our approach: with very limited
computational resources we are able to tackle a problem that is
beyond the reach of current tools even when using much more
computer power. We thus look forward to applying our method to
new and more complicated problems.

We present explicit results for the differential equations of 
the nonplanar hexa-box integrals and their solution at
symbol-level.
The differential equations are written for a pure basis of
master integrals. Most of them were available in the literature,
and we constructed the remaining ones ourselves. 
For the 5-point 
planar two-loop topologies, some pure integrals were obtained
by inserting as numerators specific minors of the Gram matrix
of the linearly-independent external and loop momenta, which
vanish in 4
dimensions. This was shown to be a good approach even for the
more complicated planar topologies, such as the double-box with 
five external legs and the penta-box, that are not
required for the particular differential equations we study in
this paper. Numerators obtained with these minors have a very
compact representation. We believe it would be very interesting
to better understand this type of numerator insertions, and why
they are good candidates for pure integrals.

Furthermore, we formulated another method for constructing 
differential equations in a very efficient manner.
This method relies on prior knowledge of the alphabet. In the 
present case, full analytic results are reconstructed from
numerical 
evaluations at 30 numerical phase-space points, which 
makes the IBP reduction trivial (even  with 
conventional Laporta-type algorithms that do not use 
unitarity-compatible IBP relations).
To make this method more generally applicable, one needs to be
able to write a conjecture for the alphabet. We note 
nevertheless that the validity of the conjecture is
verified by the success of the fitting procedure 
to reconstruct the
differential equation. We
expect that in the near future this method can be used 
in other $2$-loop $5$-point topologies for which the alphabet
has been conjectured \cite{Chicherin:2017dob}, 
but also that it will be applicable to more complicated diagrams
once techniques to construct conjectures for the alphabet become
more developed.

Our results constitute a significant step towards obtaining 
analytic expressions for all the master integrals relevant for
5-point massless amplitudes in the nonplanar sectors. Indeed, 
a convenient set of variables that rationalizes the alphabet is
known and all that remains to obtain explicit expressions in
terms of multiple polylogarithms is the calculation of boundary
conditions for the differential equations. Combined with the
recent progress in the reduction of 5-point massless amplitudes
to master integrals, we believe the
calculation of these amplitudes beyond the large-$N_c$ limit is
now well within reach.

\section*{Acknowledgment}
We thank Zvi Bern, Lance Dixon,
Fernando Febres Cordero,
Enrico Hermann, 
Kasper Larsen, Philipp Maierhoefer, Julio Parra-Martinez, and 
Yang Zhang for enlightening discussions. 
We also specially thank Harald Ita for collaboration at the 
early stages of this work and comments on the manuscript.
The work of S.A.\ and B.P.\ is supported by the Alexander von 
Humboldt Foundation, in the framework of the Sofja Kovalevskaja 
Award 2014, endowed by the German Federal Ministry of Education 
and Research.
The work of M.Z.\ is supported by the U.S. Department of Energy 
under Award Number DE-{S}C0009937.

\appendix

\section{Sector-by-sector generation of IBP relations}
\label{sec:sector}
\makeatletter{}

We generate IBP relations \emph{sector by sector}. The sector 
of an integral refers to the set of propagators that are
present in the integral, and the sector of an IBP relation
is defined by the integral (in the relation) with a maximum
number of propagators.

For the nonplanar hexa-box, there are 8 propagators, but the 
IBP relations contain integrals of lower sectors with only a 
subset of these propagators. Sectors will be arranged into 
`levels' according to the number of propagators; those with 
more propagators will be called higher sectors, and those 
with fewer propagators will be called lower sectors.

As an example, consider IBP relations for the integral topology 
shown on the top right corner of Table~\ref{tab:first1}, where 
the propagator $1/\rho_1$ of the nonplanar hexa-box has been 
canceled. Such IBP relations may be generated by 
eq.~\eqref{eq:defIBP} with
\begin{equation}
\label{eq:vEqualsRhoTimesW}
v_i^\mu = \rho_1 \, w_i^\mu \, .
\end{equation}
Eq.~\eqref{eq:GKK} is now trivially satisfied for $j=1$, so we 
only need to solve the modified equation
\begin{equation}
\label{eq:GKK-lowerSector}
w_i^\mu \rho_j = f_j \, \rho_j, \quad j=2,3, \, \dots \, . 
8 \, ,
\end{equation}
For a general sector, in the equivalent of
eq.~\eqref{eq:GKK-lowerSector} the
value of $j$ runs over all propagators that are present in 
the sector.

Of course, if we obtain a full set of solutions to the original 
equation \eqref{eq:GKK}, the solutions to 
eqs.~\eqref{eq:vEqualsRhoTimesW} and \eqref{eq:GKK-lowerSector} 
will appear as a proper subset. However, in practice we impose 
a degree bound when solving eq.~\eqref{eq:GKK} to speed up the 
computation.\footnote{Following ref.~\cite{Abreu:2017hqn}, the 
kinematic variables are absorbed into the ring variables when 
running {\tt SINGULAR}, giving a further speed-up but 
effectively making the degree bound even more stringent.} 
As a result, we 
will miss the solutions of $v_i^\mu$ related to lower sectors 
with many canceled propagators, because many inverse 
propagators, not just $\rho_1$, will have to appear in
eq.~\eqref{eq:vEqualsRhoTimesW}. Fortunately, the limited
solutions of $v_i^\mu$ still generate a full set of IBP
relations on the maximal cut of the nonplanar hexa-box, 
i.e.~ignoring integrals of lower sectors. Then we recursively
descend into lower sectors and solve the corresponding versions 
of eq.~\eqref{eq:GKK-lowerSector}, in order to generate a full 
set of IBP relations in the lower sectors. In the end, we 
combine all the relations into one linear system and solve them 
together.

In this paper, the above sector-by-sector procedure is carried 
out on cuts. For each cut, we only descend into those sectors 
where none of the cut propagators are canceled, and combining 
IBP relations from all sectors gives us a complete set of IBP 
relations on the particular cut.

\section{Momentum-twistor variables}
\label{sec:momTwist}
\makeatletter{}

Although we use the kinematic invariants in 
eq.~\eqref{eq:defSij} to compute the differential
equations, eventually one might be interested to
change to momentum-twistor 
variables \cite{Hodges:2009hk}, under which all spinor
components of external momenta have rational
components.
This ensures that the pure integrals, tabulated in Appendix 
\ref{sec:basis}, do not involve square roots in their
alphabet. 
Having a rational alphabet is also an important
step if one wants to find an explicit solution in terms of
so-called Goncharov multiple polylogarithms 
\cite{Goncharov:1998,Goncharov:2001}.

The momentum-twistor parametrization of 
external kinematics is highly redundant due to a 
$\operatorname{GL(1)}$ rescaling symmetry for each twistor and a
global $\operatorname{SL}(4)$ symmetry. 
We adopt a convenient choice of  the independent variables $x_1,
x_2, \dots , x_5$ given in  \cite{Badger:2013gxa}, but with a
cyclic shift $s_{ij}  \rightarrow s_{i+3 \, j+3}$,
\begin{align}
x_1 &= s_{45} = 1 , \nonumber\\
x_2 &= \frac{s_{45} \left(s_{51}-s_{34}\right)+s_{51} s_{12}+s_{34} s_{23}-s_{12} s_{23}-\tr_5}{2
   s_{12}}, \nonumber\\
x_3&= \frac{\left(s_{51}-s_{23}\right) \left(s_{51} s_{12}+s_{34} s_{23}-s_{12}
   s_{23}-\tr_5\right)+s_{45} \left(s_{34}-s_{51}\right) s_{51}+s_{45} \left(s_{34}+s_{51}\right)
   s_{23}}{2 \left(s_{45}+s_{51}-s_{23}\right) s_{23}}, \nonumber \\
x_4&= -\frac{s_{45} \left(s_{51}-s_{34}\right)+s_{51} s_{12}+s_{34} s_{23}-s_{12}
   s_{23}+\tr_5}{2 s_{45} \left(s_{34}-s_{51}+s_{23}\right)}, \nonumber\\
x_5&= \frac{\left(s_{51}-s_{23}\right) \left(s_{45}
   \left(s_{51}-s_{34}\right)+s_{51} s_{12}+s_{34} s_{23}-s_{12} s_{23}+\tr_5\right)}{2 s_{45} s_{51}
   \left(-s_{34}+s_{51}-s_{23}\right)},
\label{eq:twistor}
\end{align}
with $\tr_5$ defined in eq.~\eqref{eq:deftr5}.

The inverse map of eq.~\eqref{eq:twistor} is also found in
 ref.~\cite{Badger:2013gxa}, but again slightly modified by us 
 with the cyclic shift $s_{ij} \rightarrow s_{i+3 \, j+3}$,
\begin{align}\begin{split}
s_{45} &= x_1 = 1,  \\
s_{51} &= x_2 x_4,  \\
s_{12} & = \frac{ x_1 ( x_3 (x_4 - 1) + x_2 x_4 ) + x_2 x_3 (x_4 - x_5)}{x_2},  \\
s_{23} &= x_2 (x_4 - x_5),  \\
s_{34} &= -x_3 ( x_5 - 1),  \\
\tr_5 &= x_1 ( x_3 (x_4 (x_5 - 2) + 1) + x_2 x_4 (x_5 - 1) ) + x_2 x_3 (x_5 - x_4) \, .
\end{split}\end{align}
As always, we have set $s_{45}=1$ to eliminate a trivial overall scale from the problem.

\section{Table of basis integrals of uniform transcendentality}
\label{sec:basis}
\makeatletter{}

In this appendix, we explicitly write the
numerators giving pure master integrals, omitting the 
propagators of the integrals.
Some of our numerators include a propagator, which means that
the propagator will be doubled in the corresponding master
integral, see e.g.~eq.~\eqref{eq:sector01110111000}.
All nonplanar
sectors are written explicitly. For planar sectors we
write a single representative and list the ones that can be
obtained from it. Our notation for the sectors and the
corresponding numerators is given in tables \ref{tab:first1}, 
\ref{tab:first2} and \ref{tab:first3}. 
All integrals in this appendix are dimensionless if multiplied 
by $(-s_{45})^{2\epsilon}$.

Many pure nonplanar 5-point integrals are given in 
ref.~\cite{Bern:2015ple}.
The numerators are often written  in the spinor helicity
formalism. \footnote{For an introduction to 
the spinor-helicity formalism as well as various sign 
conventions adopted in this paper, we refer the readers to 
ref.~\cite{Dixon:1996wi}.}
Once manipulated to become neutral under the little group, they
can be converted to expressions in
the invariants $s_{ij}$ and the $\tr_5$ defined in 
eq.\eqref{eq:deftr5} using the following formulas,
\begin{align}\begin{split}
\langle i j \rangle [ji] &= s_{ij} , \\
\langle i j \rangle [jk] \langle kl \rangle [li] &= 2 s_{il} s_{jk}
-2 s_{ik} s_{jl} + 2 s_{ij} s_{kl} - \frac {\tr_5} {2} \operatorname{sgn} (i,j,k,l),  \\
[i j] \langle jk \rangle [kl] \langle li \rangle &= 2 s_{il} s_{jk}
-2 s_{ik} s_{jl} + 2 s_{ij} s_{kl} + \frac {\tr_5} {2} \operatorname{sgn} (i,j,k,l) \, .
\label{eq:spinorConversion}
\end{split}\end{align}
In the second and third lines of the equation above, 
$(i, j, k, l)$ are  
permutations of $(1,2,3,4)$, and the $\operatorname{sgn}$ 
symbol denotes the signature of the permutation. When dealing 
with expressions involving $|5 \rangle$ or $|5]$, we 
first rewrite the expressions using only the spinors of the 
legs $1,2,3,4$, and then apply the last two relations of 
eq.~\eqref{eq:spinorConversion}. After rewriting the numerators 
in terms of $s_{ij}$ and $\tr_5$, we define the even-parity and 
odd-parity parts of any numerator $\mathcal N$ as
\begin{align}\begin{split}
\text{even part of } \mathcal N &= \frac 1 2 \big(\mathcal N +
\mathcal N|_{\tr_5 \rightarrow - \tr_5} \big), \\
\text{odd part of } \mathcal N &= \frac 1 2 \big(\mathcal N -
\mathcal N|_{\tr_5 \rightarrow - \tr_5} \big) \, .
\end{split}\end{align}
Recall that $\tr_5$ is imaginary in Minkowskian signature, and 
is proportional to a contraction with the Levi-Civita 
$\varepsilon$-tensor, which is odd
under parity.
By projecting the numerators of ref.~\cite{Bern:2015ple}
onto even and odd parts to construct our basis, 
we make sure that the 
integrands have definite parity, and are either purely real or
purely imaginary. Nontrivial complex values only arise after 
loop integration.

\begin{itemize}

\item
Sector $F[1,1,1,1, 1,1,1,1, 0,0,0]$: pure numerators from
\cite{Bern:2015ple},
\begin{align}\begin{split}
\mathcal N_1 &=  2 \times \text{odd part of }  \\
&\qquad \left \{ [13] \left( l_1 + \frac{P_{45} \cdot \tilde \lambda_3 \tilde \lambda_1 }{[13]} \right)^2 \langle 15 \rangle [54] \langle 43 \rangle \left( l_1 + k_4 \right)^2 \right \} \, , \\
\mathcal N_2 &= \mathcal N_1 \big|_{4 \leftrightarrow 5} \, , \\
\mathcal N_3 &= s_{12} s_{23} \left[ \left( l_1 + p_4 \right)^2 \left( l_1 + p_5 \right)^2 - l_1^2 \left( l_1 + p_4 + p_5 \right)^2 \right] \, .
\label{eq:pureIntHexabox}
\end{split}\end{align}
In eq.~\eqref{eq:pureIntHexabox}, 
$P_{45} \cdot \tilde \lambda_3 \tilde
\lambda_1$ denotes a Lorentz 4-vector given by:
\begin{equation}
P_{45} \cdot \tilde \lambda_3 \tilde
\lambda_1=
\frac 1 2 [1 | \gamma^\mu  (\slashed p_4 + \slashed p_5) | 3] \, .
\end{equation}

\item
Sector $F[1,1,1,0, 1,1,1,1, 0,0,0]$: pure numerators from \cite{Bern:2015ple},
\begin{align}\begin{split}
\mathcal N_{11} &= 2 \times \text{even part of } \left \{ \langle 15 \rangle [45] \langle 24 \rangle [12] \left( l_1 - \frac{\langle 12 \rangle} {\langle 42 \rangle} \lambda_4 \tilde \lambda_1 \right)^2 \right\} \, , \label{eq:pureN11} \\
\mathcal N_{12} &= 2 \times \text{odd part of } \Big \{ s_{12} \langle 14 \rangle [15] \langle 5 | l_1 | 4 ] \Big \} \, , \\
\mathcal N_{13} &= 2 \times \text{even part of } \Big \{ s_{12} \langle 14 \rangle [15] \langle 5 | l_1 | 4 ] \Big \} \, .
\end{split}\end{align}
In eq.~\eqref{eq:pureN11}, $\lambda_4 \tilde \lambda_1$ denotes 
a Lorentz 4-vector given by:
\begin{equation}
\lambda_4 \tilde \lambda_1=
\frac 1 2 [1 | \gamma^\mu | 4 \rangle
\, .
\end{equation}
\item
Sector $F[0,1,1,1, 1,1,1,1, 0,0,0]$: related to the previous sector by a diagram symmetry, and the pure numerator are,
\begin{align}\begin{split}
\mathcal N_{4} &= 2 \times \text{even part of } \left \{ \langle 35 \rangle [45] \langle 24 \rangle [32] \left( l_{\rho_4} - \frac{\langle 32 \rangle} {\langle 42 \rangle} \lambda_4 \tilde \lambda_3 \right)^2 \right\} \, , \\
\mathcal N_{5} &= (-2) \times \text{odd part of } \Big \{ s_{23} \langle 34 \rangle [35] \langle 5 | l_{\rho_4} | 4 ] \Big \} \, , \\
\mathcal N_{6} &= 2 \times \text{even part of } \Big \{ s_{23} \langle 34 \rangle [35] \langle 5 | l_{\rho_4} | 4 ] \Big \} \, ,
\end{split}\end{align}
where $l_{\rho_4}=-l_1+p_{1}+p_{2}+p_{3}$.
\item
Sector $F[1,0,1,1,1,1,1,1,0,0,0]$: Pure numerator from
\cite{Bern:2015ple},
\begin{align}\begin{split}
\mathcal N_{7} &= s_{35}s_{45}(l_{\rho_4}+p_4)^2,\\
\mathcal N_{8} &=s_{34}s_{45}(l_{\rho_4}+p_5)^2,
\end{split}\end{align}
where $l_{\rho_4}=-l_1+p_{1}+p_{2}+p_{3}$.
\item
Sector $F[1,1,0,1,1,1,1,1,0,0,0]$: Numerators for 
this sector can be constructed from the previous one to get 
\begin{align}\begin{split}
\mathcal N_{9} &= s_{14}s_{45}(l_1+p_5)^2,\\
\mathcal N_{10}&= s_{51}s_{45}(l_1+p_4)^2.
\end{split}\end{align}
\item
Sector $F[0,1,0,1,1,1,1,1,0,0,0]$: Pure numerators from
\cite{Gehrmann:2001ck},
\begin{align}\begin{split}
\mathcal N_{14} &=2(s_{51}+s_{45})\,
l_{\rho_4}\cdot p_4,\\
\mathcal N_{15} &=s_{23} s_{45}\,,
\label{eq:sec1415}
\end{split}\end{align}
where $l_{\rho_4}=-l_1+p_{1}+p_{2}+p_{3}$.
\item
Sector $F[0,1,1,0, 1,1,1,1, 0,0,0]$: Pure numerator from
\cite{Bern:2015ple},
\begin{equation}
\mathcal N_{16} = \langle 2 | 3 | 4] \langle 4 | 5 | 2 ] -
\langle 4 | 3 | 2] \langle 2 | 5 | 4 ] = - \tr_5 \, .
\end{equation}
\item
Sector $F[0,1,1,1,0,1,1,1,0,0,0]$:
\begin{align}\begin{split}
\mathcal N_{17} &=\frac{\tr_5}{2(4-d)}\frac{\mu_{11}}{\rho_6}, \label{eq:sector01110111000} \\
\mathcal N_{18} &=s_{23}(s_{12}-s_{34}).
\end{split}\end{align}
The numerators $\mathcal N_{i}$ for $i\in\{19,24,26\}$ are
obtained from $\mathcal N_{17}$  by analogy.
The numerators
$\mathcal N_{i}$ for $i\in\{20,25,27\}$ are
obtained from $\mathcal N_{18}$ by analogy.
Since we always factor out the propagators (each with a single 
power), an additional power of $1 / \rho_6$ in 
eq.~\eqref{eq:sector01110111000} means a doubled propagator 
$1 / \rho_6^2$ in the integral.

\item
Sector $F[1,0,0,1,1,1,1,1,0,0,0]$: Pure numerator from
\cite{Gehrmann:2001ck},
\begin{equation}
\mathcal N_{21} =s_{45}^2.
\end{equation}
\item
Sector $F[1,0,1,0,1,1,1,1,0,0,0]$: The numerators for this 
sector are obtained in the same way as those of 
eq.~\eqref{eq:sec1415},
\begin{align}\begin{split}
\mathcal N_{22} &=2 (s_{35}+s_{45})\,l_1\cdot p_4
=2 (s_{12}-s_{34})\,l_1\cdot p_4,\\
\mathcal N_{23} &=s_{12} s_{45}.
\end{split}\end{align}
\item
Sector $F[1,1,1,1,0,1,0,1,0,0,0]$:
\begin{align}\begin{split}
\mathcal N_{28} &=\frac{\tr_5}{2(4-d)}\frac{\mu_{11}}{\rho_6},\\
\mathcal N_{29} &=\frac{2(d-3)s_{12}s_{23}}{4-d}.
\end{split}\end{align}
The numerator $\mathcal N_{30}$ is
obtained from $\mathcal N_{28}$  by analogy.
The numerator
$\mathcal N_{31}$ is
obtained from $\mathcal N_{29}$  by analogy.
\item
Sector $F[0,0,1,0,1,1,1,1,0,0,0]$:
\begin{equation}
\mathcal N_{32} =s_{34}+s_{35}-s_{12}=-s_{45}.
\end{equation}
The numerators $\mathcal N_{i}$ for 
$i\in\{33,34,35,40,41,42,43,53,54\}$ are obtained by analogy.
\item
Sector $F[0,1,0,1,0,1,1,1,0,0,0]$:
\begin{align}\begin{split}
\mathcal N_{36} &=\frac{2s_{45}s_{51}}{4-d}\frac{1}{\rho_6},\\
\mathcal N_{37} &=s_{45}+s_{51}.
\end{split}\end{align}
The numerators $\mathcal N_{i}$ for $i\in\{39,48,50\}$ are
obtained from $\mathcal N_{37}$  by analogy. The numerators
$\mathcal N_{i}$ for $i\in\{38,47,49\}$ are
obtained from $\mathcal N_{36}$  by analogy.
\item
Sector $F[0,1,1,1,0,1,0,1,0,0,0]$:
\begin{equation}
\mathcal N_{44} =\frac{(d-3)s_{23}}{d-4}.
\end{equation}
The numerators $\mathcal N_{i}$ for $i\in\{45, 46, 57, 58, 59\}$
are obtained by analogy.
\item
Sector $F[1,0,1,1,0,1,0,1,0,0,0]$:
\begin{equation}
\mathcal N_{51} =\frac{(d-3)(s_{45}-s_{12})}{d-4}.
\end{equation}
The numerators $\mathcal N_{i}$ for $i\in\{52,55,56\}$
are obtained by analogy.
\item
Sector $F[0,1,0,1,0,1,0,1,0,0,0]$:
\begin{equation}
\mathcal N_{60} = \frac{(3d-10)(d-3)}{(d-4)^2}.
\end{equation}
The numerators $\mathcal N_{i}$ for $i\in\{61,\ldots,66\}$
are obtained by analogy. Note that sector 63 
has a single external mass, but given that these integrals are
dimensionless the normalization does not change. 
\item
Sector $F[0,0,0,1,0,1,1,0,0,0,0]$:
\begin{equation}\label{eq:sunriseNorm}
\mathcal N_{67} = \frac{(3d-8)(3d-10)(d-3)}{(d-4)^3s_{45}}.
\end{equation}
The numerators $\mathcal N_{i}$ for $i\in\{68,\ldots,73\}$
are obtained by analogy.

\end{itemize}

\begin{table}[h]
\centering
 \begin{tabular}{c |  c || c  | c } 
 Sector and diagram & Numerators &
 Sector and diagram & Numerators \\ \hline
 \includegraphics[width=2.5cm]{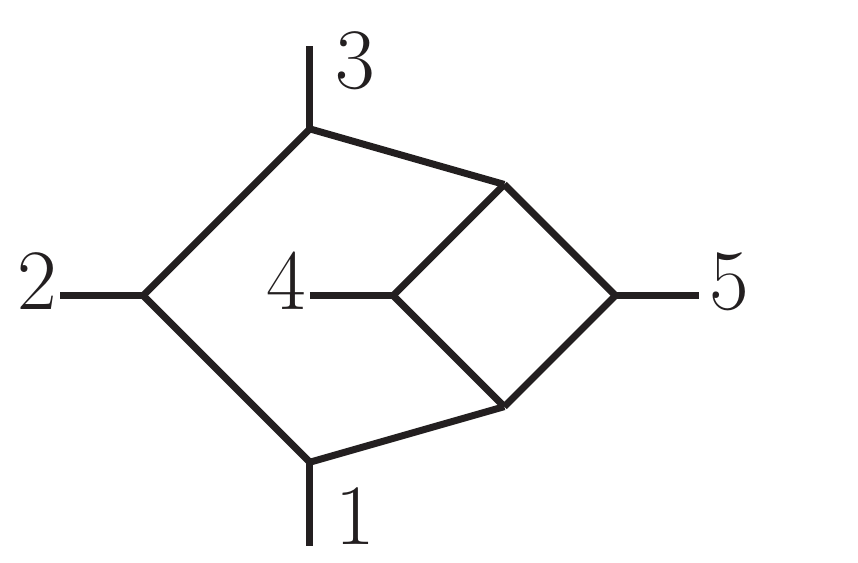} &
 \raisebox{.4cm}{$\mathcal{N}_{1}$, 
 $\mathcal{N}_{2}$, 
 $\mathcal{N}_{3}$}
  & 
 \raisebox{.2cm}{\includegraphics[width=2.5cm]
 {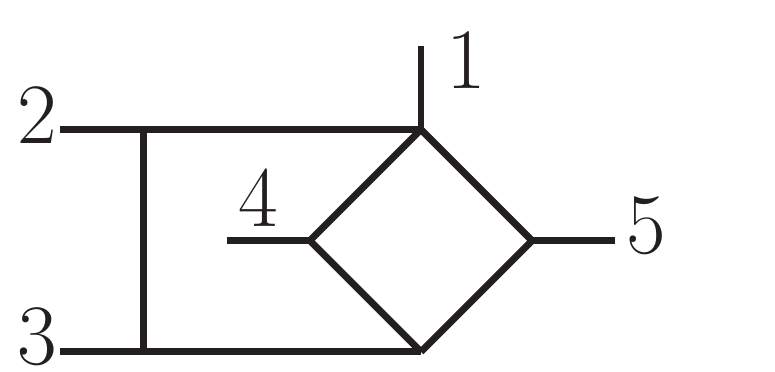}} & 
 \raisebox{.4cm}{$\mathcal{N}_{4}$, 
 $\mathcal{N}_{5}$, 
 $\mathcal{N}_{6}$} \\
  \footnotesize{$F[1,1,1,1, 1,1,1,1, 0,0,0]$} 
  & &
 \footnotesize{$F[0,1,1,1, 1,1,1,1, 0,0,0]$} &  \\\hline
 \includegraphics[width=2.5cm]{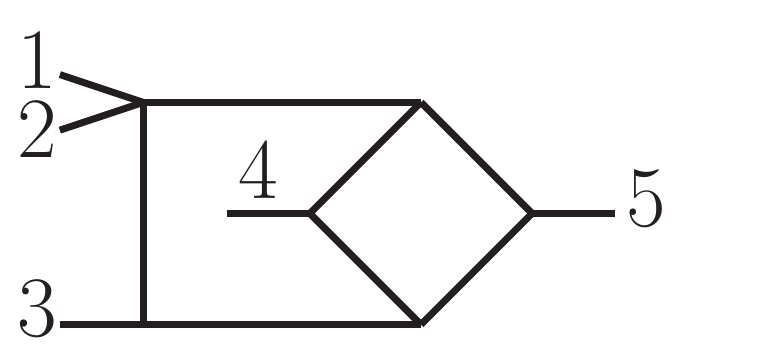} & 
 \raisebox{.2cm}{$\mathcal{N}_{7}$, 
 $\mathcal{N}_{8}$}&
 \includegraphics[width=2.5cm]{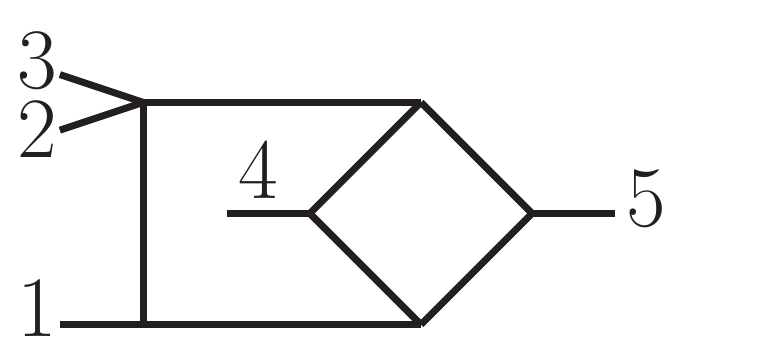} & 
 \raisebox{.2cm}{$\mathcal{N}_{9}$, 
 $\mathcal{N}_{10}$} \\
 \footnotesize{$F[1,0,1,1, 1,1,1,1, 0,0,0]$} 
  & &
 \footnotesize{$F[1,1,0,1, 1,1,1,1, 0,0,0]$}  & \\\hline
 \includegraphics[width=2.5cm]{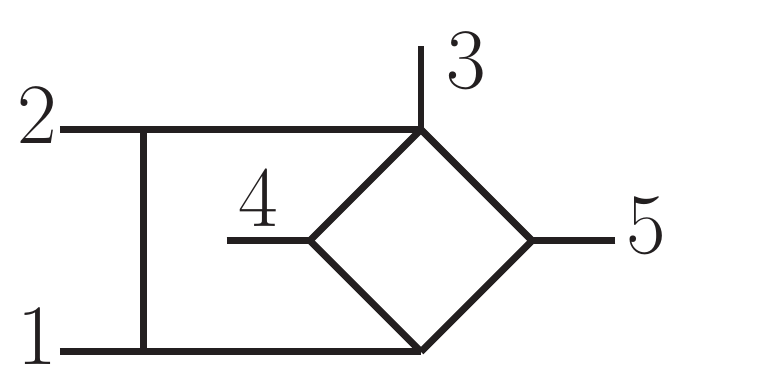} &
 \raisebox{.2cm}{$\mathcal{N}_{11}$, 
 $\mathcal{N}_{12}$, 
 $\mathcal{N}_{13}$} &
 \includegraphics[width=2.5cm]{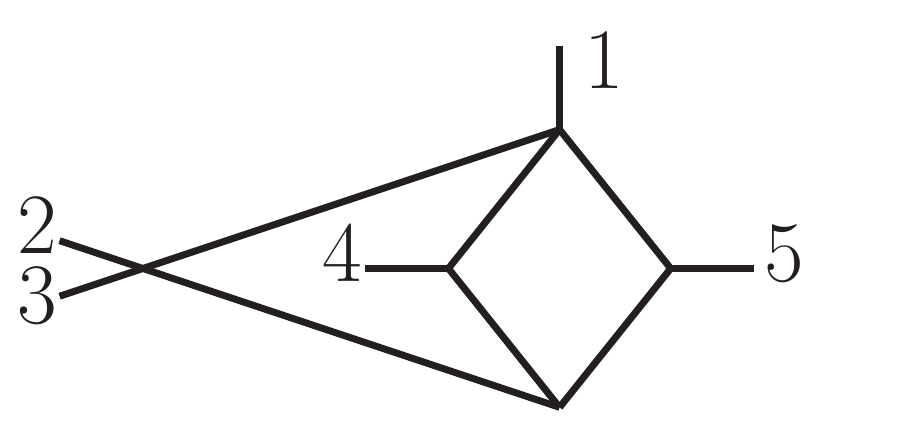} & 
 \raisebox{.2cm}{$\mathcal{N}_{14}$, 
 $\mathcal{N}_{15}$} \\
 \footnotesize{$F[1,1,1,0, 1,1,1,1, 0,0,0]$} 
  & &
 \footnotesize{$F[0,1,0,1, 1,1,1,1, 0,0,0]$}  & \\\hline
 \includegraphics[width=2.5cm]{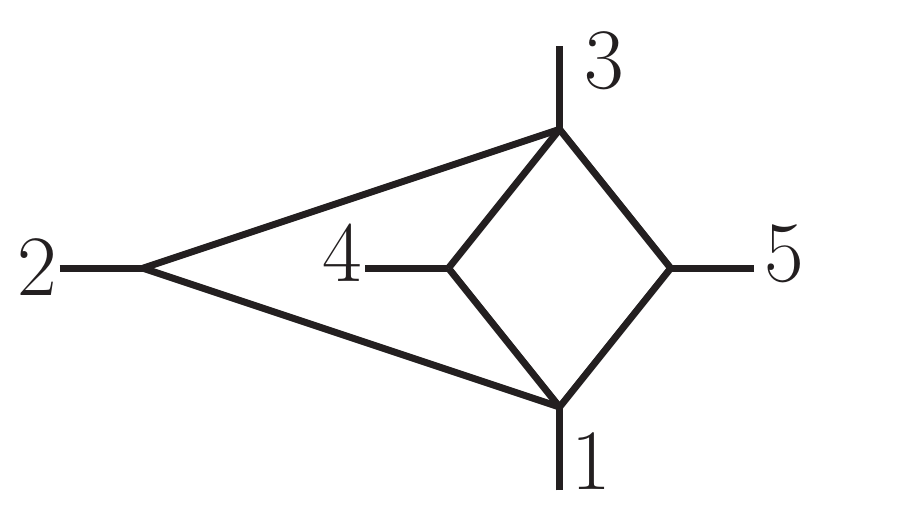}  &
 \raisebox{.4cm}{$\mathcal{N}_{16}$} &
 \includegraphics[width=2.5cm]{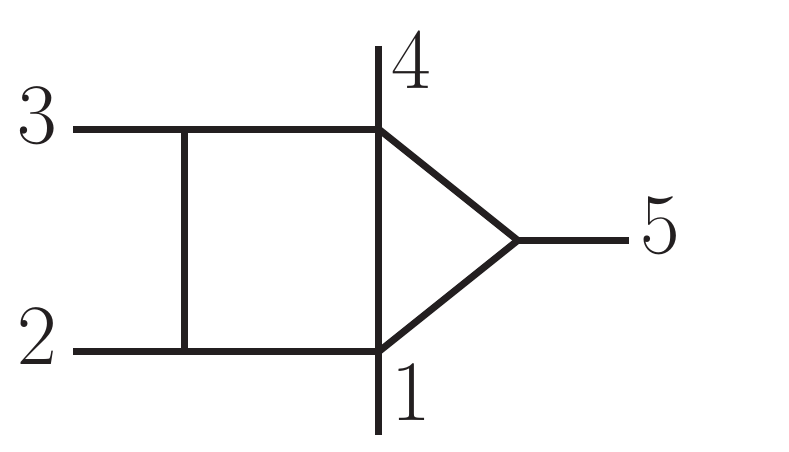} &
 \raisebox{.4cm}{$\mathcal{N}_{17}$, 
 $\mathcal{N}_{18}$}\\
 \footnotesize{$F[0,1,1,0, 1,1,1,1, 0,0,0]$} 
  & &
 \footnotesize{$F[0,1,1,1, 0,1,1,1, 0,0,0]$}  & \\\hline
 \includegraphics[width=2.5cm]{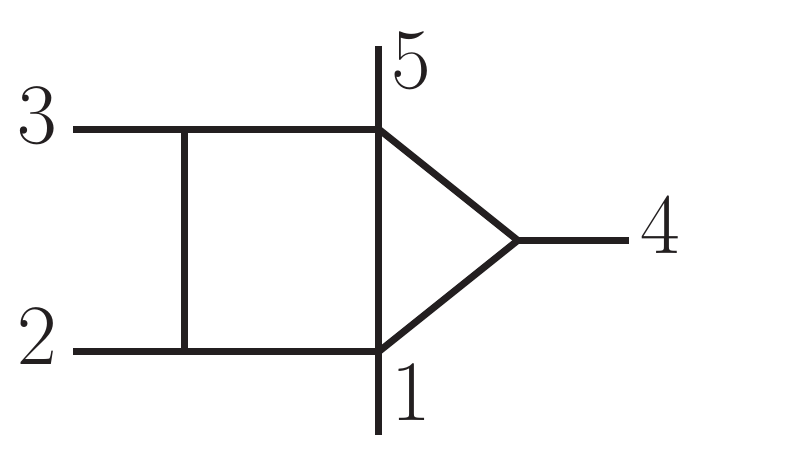}  &
 \raisebox{.4cm}{$\mathcal{N}_{19}$, 
 $\mathcal{N}_{20}$} &
 \raisebox{.25cm}{\includegraphics[width=2.5cm]
 {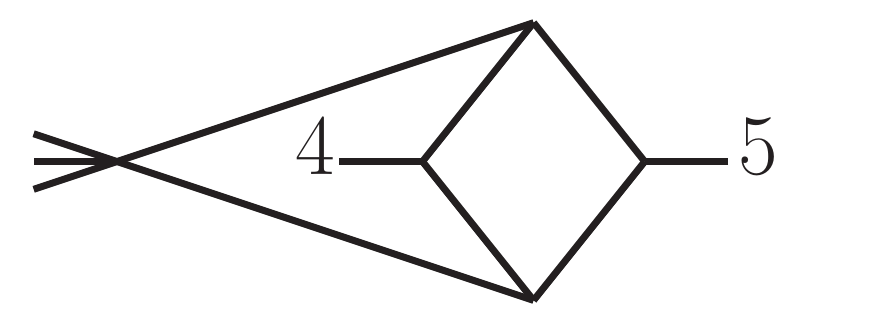}}&
 \raisebox{.4cm}{$\mathcal{N}_{21}$}\\
 \footnotesize{$F[0,1,1,1, 1,1,1,0, 0,0,0]$} 
  & &
 \footnotesize{$F[1,0,0,1, 1,1,1,1, 0,0,0]$}  & \\\hline
 \raisebox{.25cm}{\includegraphics[width=2.5cm]
 {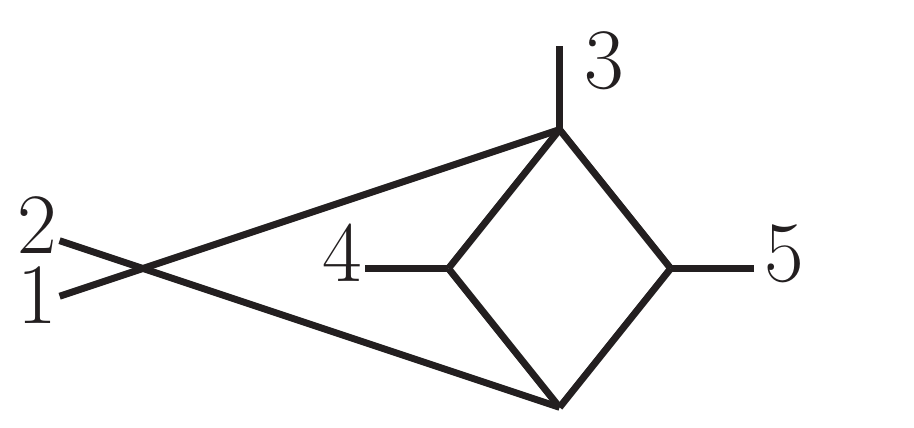}} &
 \raisebox{.4cm}{$\mathcal{N}_{22}$, 
 $\mathcal{N}_{23}$} &
 \raisebox{.1cm}{\includegraphics[width=2.5cm]
 {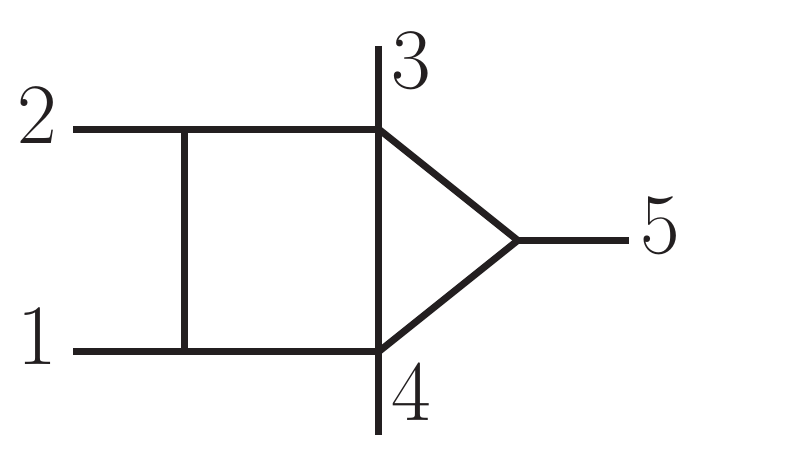}} &
 \raisebox{.4cm}{$\mathcal{N}_{24}$, 
 $\mathcal{N}_{25}$}\\
 \footnotesize{$F[1,0,1,0, 1,1,1,1, 0,0,0]$} 
  & &
 \footnotesize{$F[1,1,1,0, 1,0,1,1, 0,0,0]$}  & \\\hline
 \raisebox{.25cm}{\includegraphics[width=2.5cm]
 {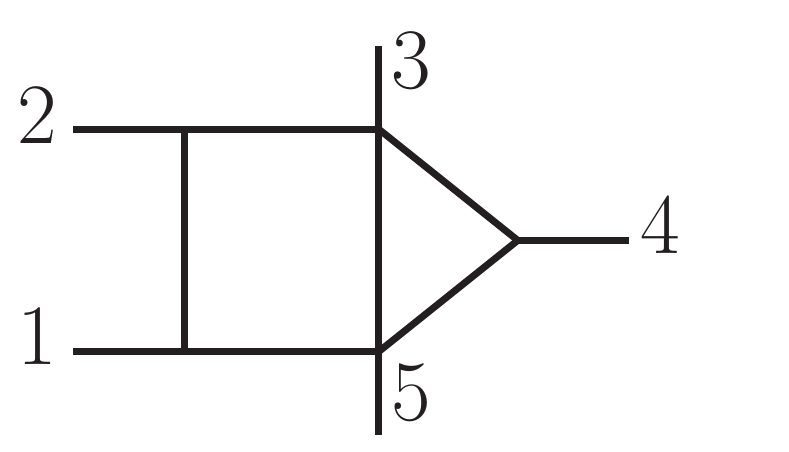}}  &
 \raisebox{.6cm}{$\mathcal{N}_{26}$, 
 $\mathcal{N}_{27}$} &
 \raisebox{.1cm}{\includegraphics[width=2.2cm]
 {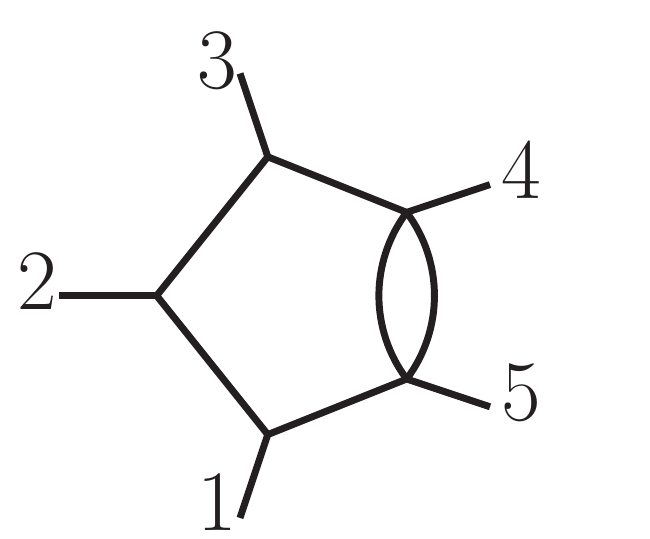}} & 
 \raisebox{.6cm}{$\mathcal{N}_{28}$, 
 $\mathcal{N}_{29}$}\\
 \footnotesize{$F[1,1,1,0, 1,1,0,1, 0,0,0]$} 
  & &
 \footnotesize{$F[1,1,1,1, 0,1,0,1, 0,0,0]$}  & \\\hline
 \raisebox{.1cm}{\includegraphics[width=2.2cm]
 {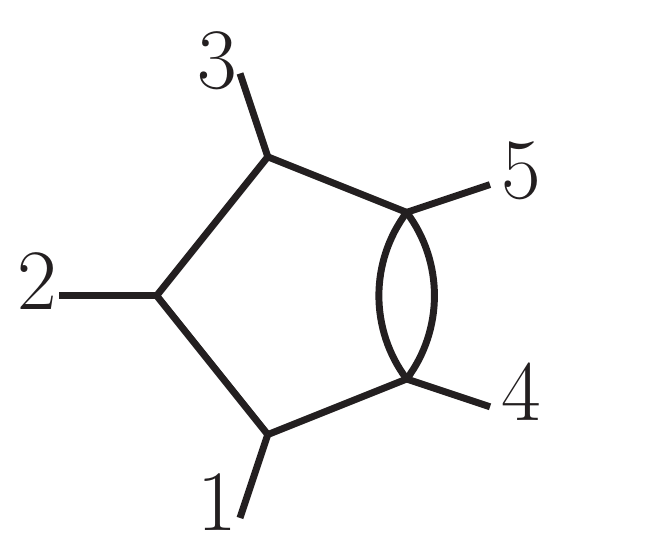}} &
 \raisebox{.6cm}{$\mathcal{N}_{30}$, 
 $\mathcal{N}_{31}$} & 
 \raisebox{.4cm}{\includegraphics[width=2.5cm]
 {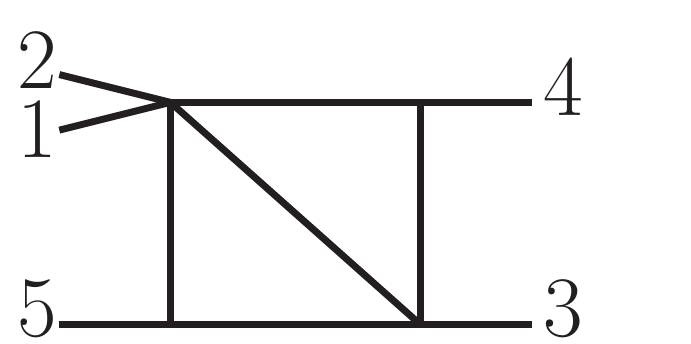}} & 
 \raisebox{.6cm}{$\mathcal{N}_{32}$} \\
 \footnotesize{$F[1,1,1,1, 1,0,1,0, 0,0,0]$} 
  & &
 \footnotesize{$F[0,0,1,0, 1,1,1,1, 0,0,0]$} & \\\hline
  \includegraphics[width=2.5cm]{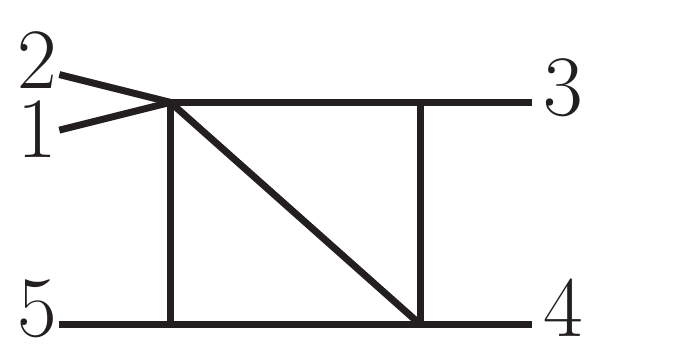}  &
 \raisebox{.2cm}{$\mathcal{N}_{33}$} &
 \includegraphics[width=2.5cm]{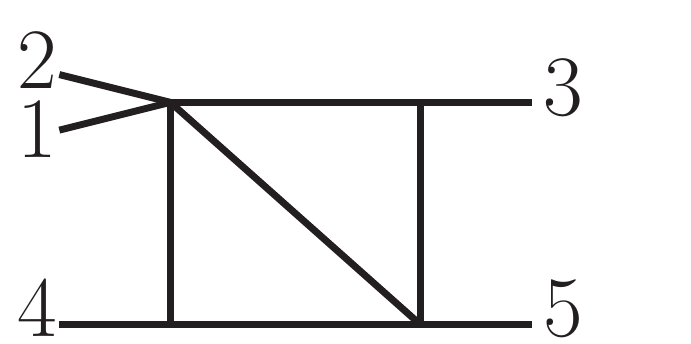}  &
  \raisebox{.2cm}{$\mathcal{N}_{34}$}\\
 \footnotesize{$F[0,0,1,1, 0,1,1,1, 0,0,0]$} 
  & &
 \footnotesize{$F[0,0,1,1, 1,1,1,0, 0,0,0]$}  & \\\hline
 \includegraphics[width=2.5cm]{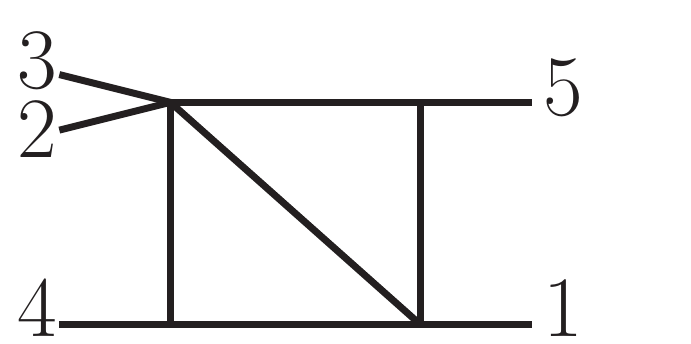}  &
  \raisebox{.2cm}{$\mathcal{N}_{35}$} &
 \includegraphics[width=2.5cm]{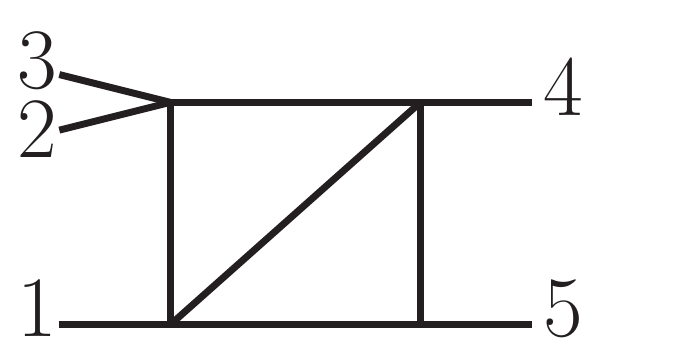}  &
 \raisebox{.2cm}{$\mathcal{N}_{36}$, 
 $\mathcal{N}_{37}$}  \\
 \footnotesize{$F[0,1,0,0, 1,1,1,1, 0,0,0]$} 
  & &
 \footnotesize{$F[0,1,0,1, 0,1,1,1, 0,0,0]$} & 
 \end{tabular}
 \caption{Diagrammatic representation of each sector and
 associated numerators.}
 \label{tab:first1}
\end{table}

\begin{table}[]
\centering
 \begin{tabular}{c | c  || c | c  } 
 Sector and diagram & Numerators &
 Sector and diagram & Numerators \\ \hline
 \includegraphics[width=2.5cm]{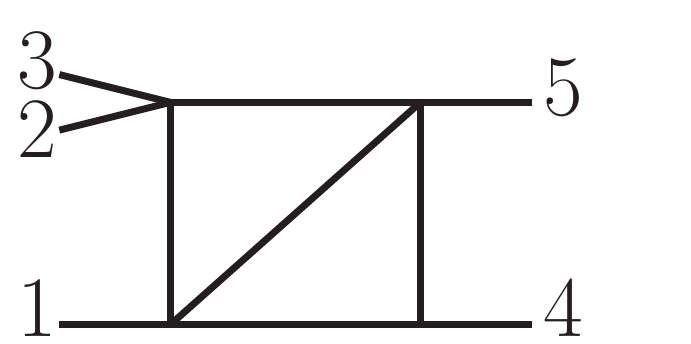}  &
 \raisebox{.2cm}{$\mathcal{N}_{38}$, 
 $\mathcal{N}_{39}$} &
 \includegraphics[width=2.5cm]{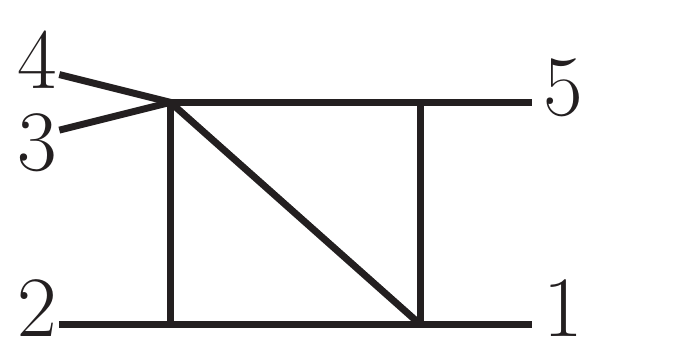}  &
  \raisebox{.2cm}{$\mathcal{N}_{40}$} \\
 \footnotesize{$F[0,1,0,1, 1,1,1,0, 0,0,0]$} 
  & &
 \footnotesize{$F[0,1,1,0, 0,1,1,1, 0,0,0]$}  & \\\hline
 \includegraphics[width=2.5cm]{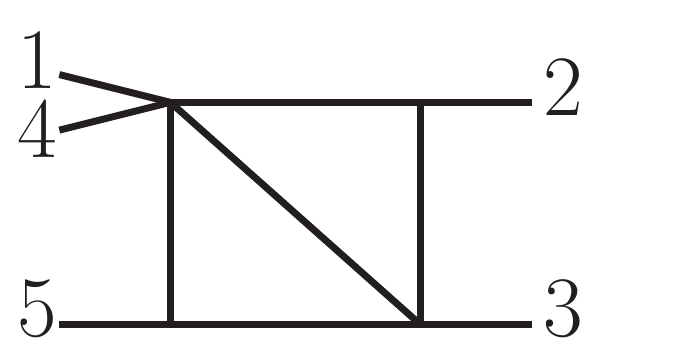}  &
  \raisebox{.2cm}{$\mathcal{N}_{41}$} &
 \includegraphics[width=2.5cm]{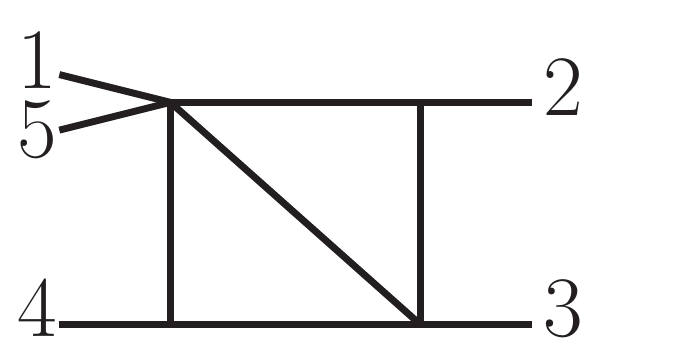}  &
  \raisebox{.2cm}{$\mathcal{N}_{42}$}\\
 \footnotesize{$F[0,1,1,0, 1,0,1,1, 0,0,0]$} 
  & &
 \footnotesize{$F[0,1,1,0, 1,1,0,1, 0,0,0]$}  & \\\hline
 \includegraphics[width=2.5cm]{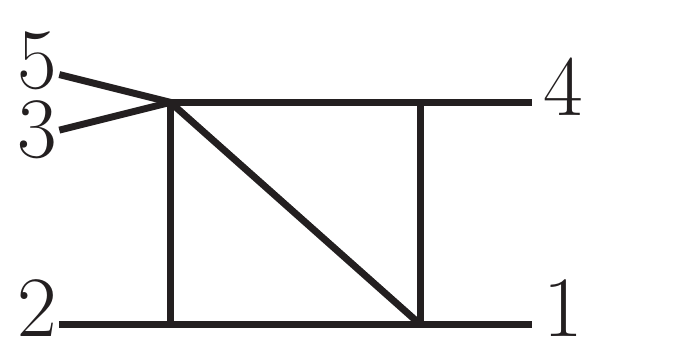}  &
  \raisebox{.2cm}{$\mathcal{N}_{43}$} &
 \includegraphics[width=2.5cm]{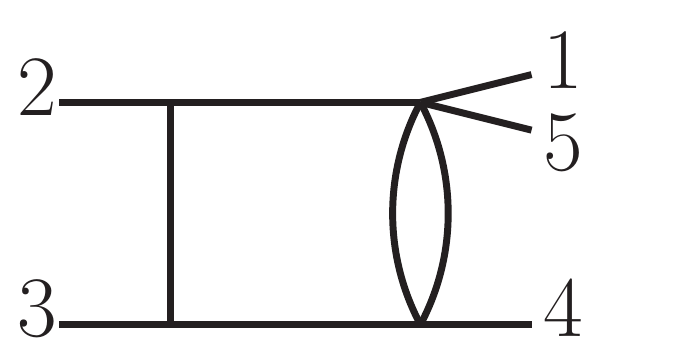}  &
  \raisebox{.2cm}{$\mathcal{N}_{44}$} \\ 
 \footnotesize{$F[0,1,1,0, 1,1,1,0, 0,0,0]$} 
  & &
 \footnotesize{$F[0,1,1,1, 0,1,0,1, 0,0,0]$}  & \\\hline
 \includegraphics[width=2.5cm]{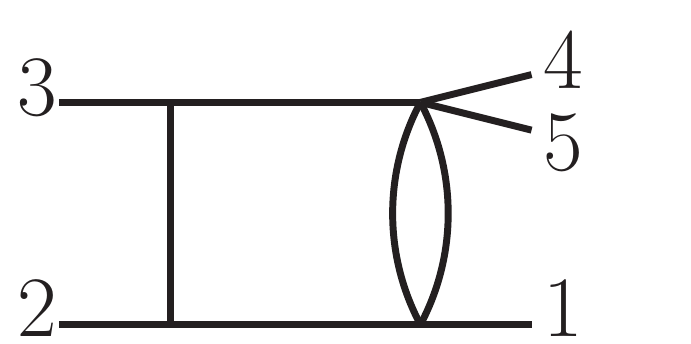}  &
  \raisebox{.2cm}{$\mathcal{N}_{45}$} &
 \includegraphics[width=2.5cm]{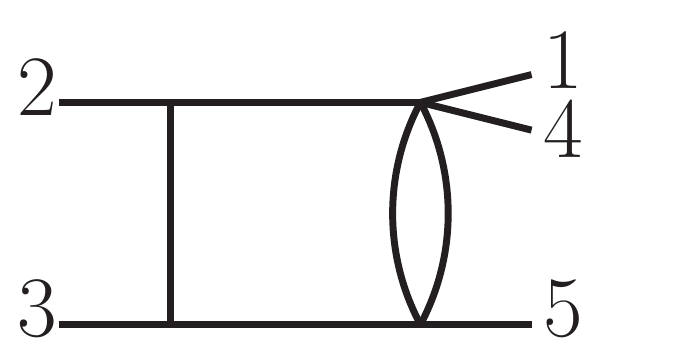} &
  \raisebox{.2cm}{$\mathcal{N}_{46}$}\\
 \footnotesize{$F[0,1,1,1, 0,1,1,0, 0,0,0]$} 
  & &
 \footnotesize{$F[0,1,1,1, 1,0,1,0, 0,0,0]$}  & \\\hline
 \includegraphics[width=2.5cm]{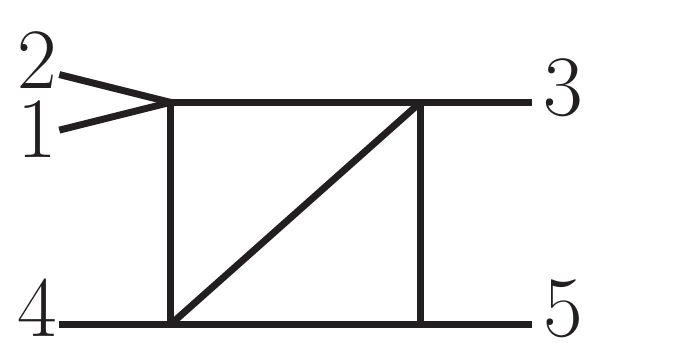}  &
 \raisebox{.2cm}{$\mathcal{N}_{47}$, 
 $\mathcal{N}_{48}$}  &
 \includegraphics[width=2.5cm]{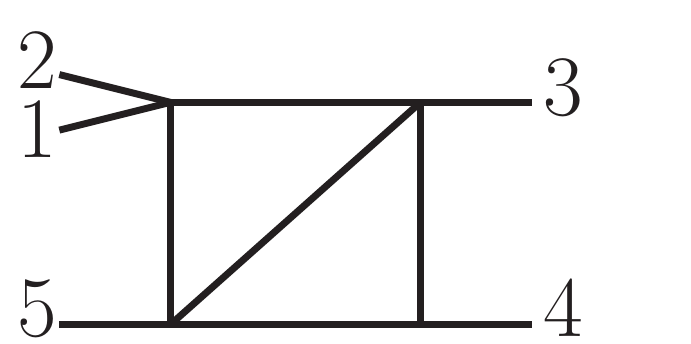}  &
 \raisebox{.2cm}{$\mathcal{N}_{49}$, 
 $\mathcal{N}_{50}$}  \\
 \footnotesize{$F[1,0,1,0, 1,0,1,1, 0,0,0]$} 
  & &
 \footnotesize{$F[1,0,1,0, 1,1,0,1, 0,0,0]$}  & \\\hline
 \includegraphics[width=2.5cm]{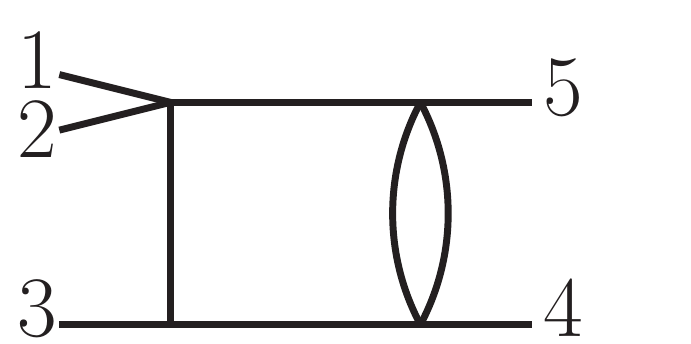}  &
  \raisebox{.2cm}{$\mathcal{N}_{51}$} &
 \includegraphics[width=2.5cm]{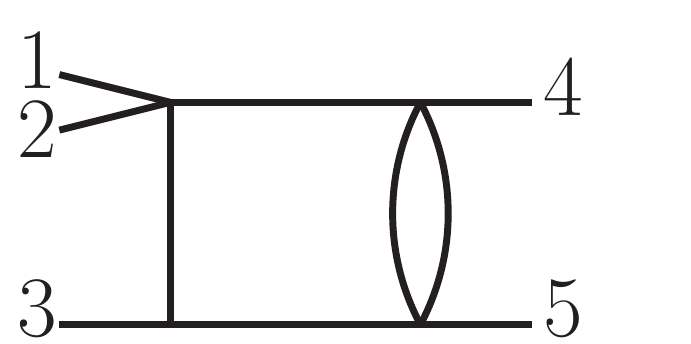}  &
  \raisebox{.2cm}{$\mathcal{N}_{52}$} \\
 \footnotesize{$F[1,0,1,1, 0,1,0,1, 0,0,0]$} 
  & &
 \footnotesize{$F[1,0,1,1, 1,0,1,0, 0,0,0]$}  & \\\hline
 \includegraphics[width=2.5cm]{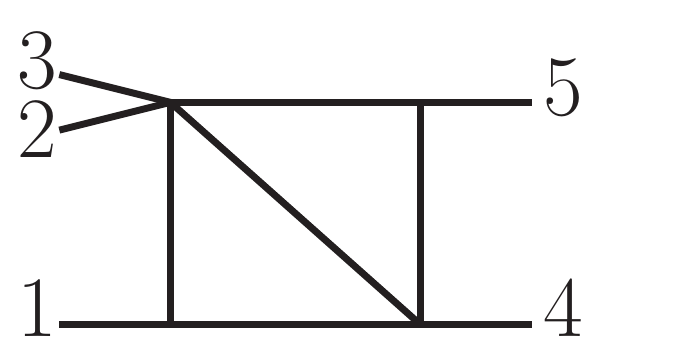}  &
  \raisebox{.2cm}{$\mathcal{N}_{53}$} &
 \includegraphics[width=2.5cm]{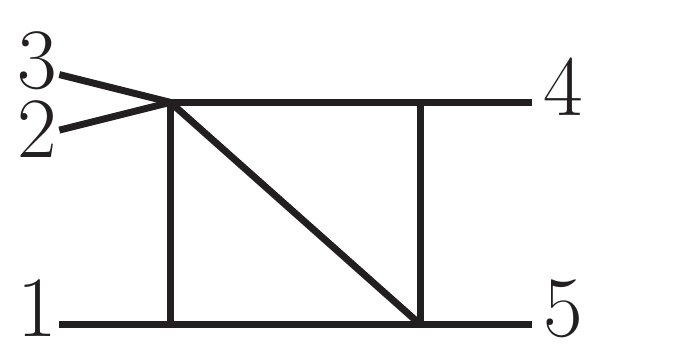}  &
  \raisebox{.2cm}{$\mathcal{N}_{54}$} \\
 \footnotesize{$F[1,1,0,0, 1,0,1,1, 0,0,0]$} 
  & &
 \footnotesize{$F[1,1,0,0, 1,1,0,1, 0,0,0]$}  & \\\hline
 \includegraphics[width=2.5cm]{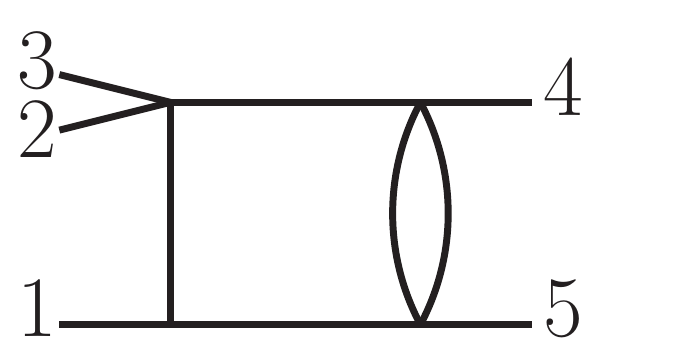}  &
  \raisebox{.2cm}{$\mathcal{N}_{55}$} &
 \includegraphics[width=2.5cm]{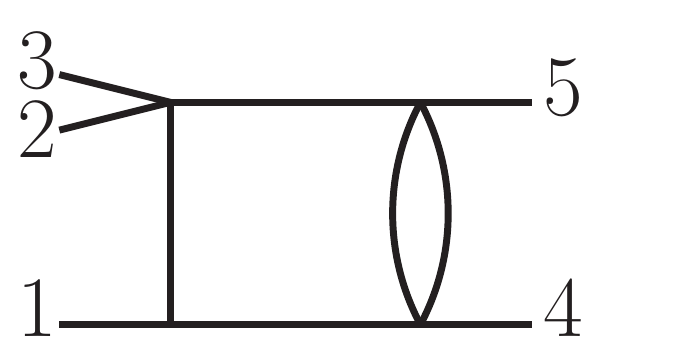}  &
  \raisebox{.2cm}{$\mathcal{N}_{56}$} \\
 \footnotesize{$F[1,1,0,1, 0,1,0,1, 0,0,0]$} 
  & &
 \footnotesize{$F[1,1,0,1, 1,0,1,0, 0,0,0]$} & \\\hline
 \includegraphics[width=2.5cm]{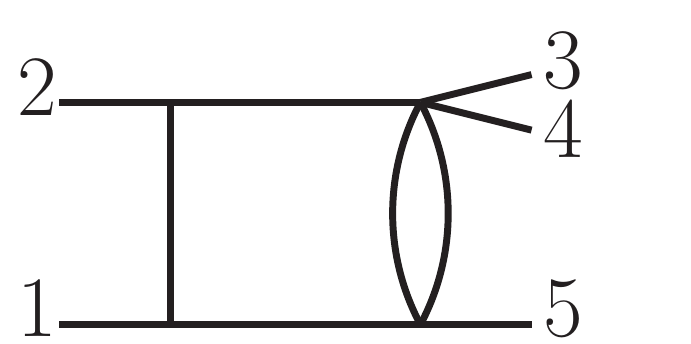}  &
 \raisebox{.2cm}{$\mathcal{N}_{57}$} &
 \includegraphics[width=2.5cm]{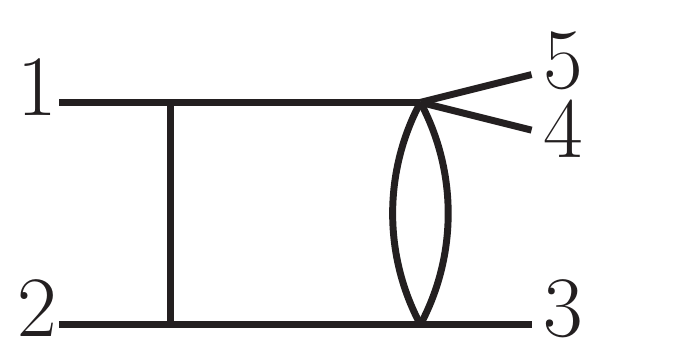}  &
  \raisebox{.2cm}{$\mathcal{N}_{58}$} \\
 \footnotesize{$F[1,1,1,0, 0,1,0,1, 0,0,0]$} 
  & &
 \footnotesize{$F[1,1,1,0, 1,0,0,1, 0,0,0]$}  & \\\hline
\includegraphics[width=2.5cm]{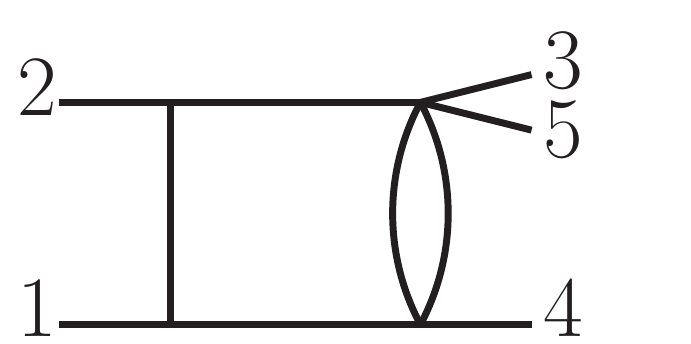} &
 \raisebox{.2cm}{$\mathcal{N}_{59}$} &
 \includegraphics[width=2.5cm]{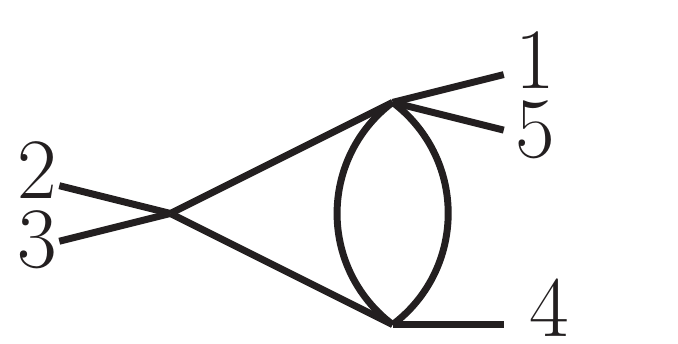} &
  \raisebox{.2cm}{$\mathcal{N}_{60}$}  \\
 \footnotesize{$F[1,1,1,0, 1,0,1,0, 0,0,0]$}
  & &
 \footnotesize{$F[0,1,0,1, 0,1,0,1, 0,0,0]$} & 
 \end{tabular}
 \caption{Diagrammatic representation of each sector and
 associated numerators.}
 \label{tab:first2}
\end{table}

\begin{table}[]
\centering
 \begin{tabular}{c | c  || c | c  } 
 Sector and diagram & Numerators &
 Sector and diagram & Numerators \\ \hline

 \includegraphics[width=2.5cm]{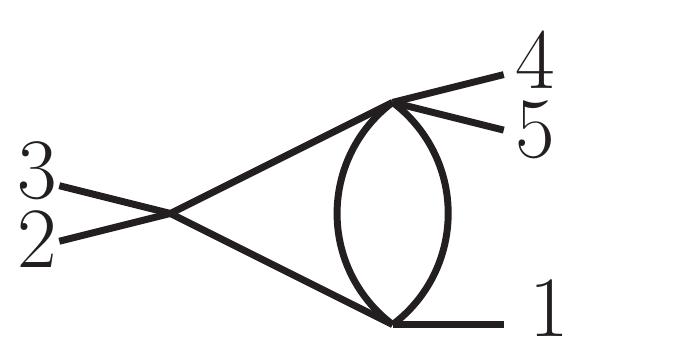}  &
  \raisebox{.2cm}{$\mathcal{N}_{61}$} &
 \includegraphics[width=2.5cm]{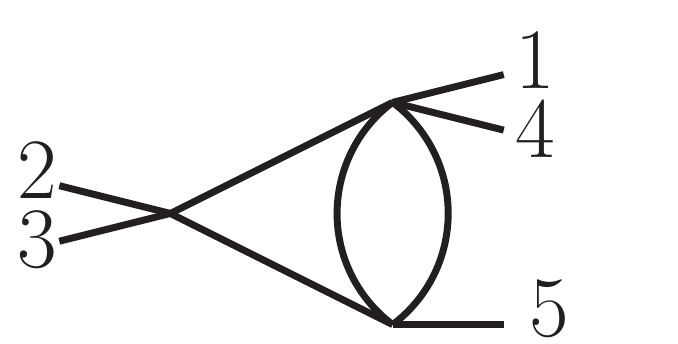} &
 \raisebox{.2cm}{$\mathcal{N}_{62}$}   \\
 \footnotesize{$F[0,1,0,1, 0,1,1,0, 0,0,0]$}
  & &
 \footnotesize{$F[0,1,0,1, 1,0,1,0, 0,0,0]$} &  \\\hline
 \includegraphics[width=2.5cm]{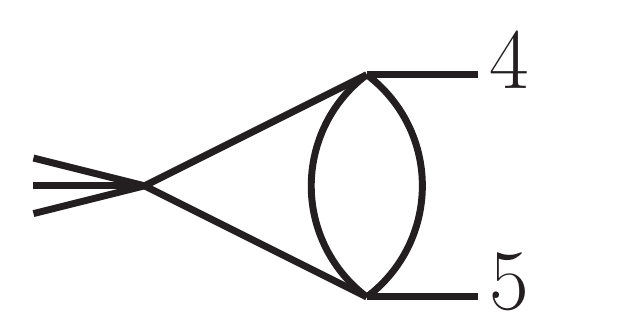} &
  \raisebox{.2cm}{$\mathcal{N}_{63}$} &
 \includegraphics[width=2.5cm]{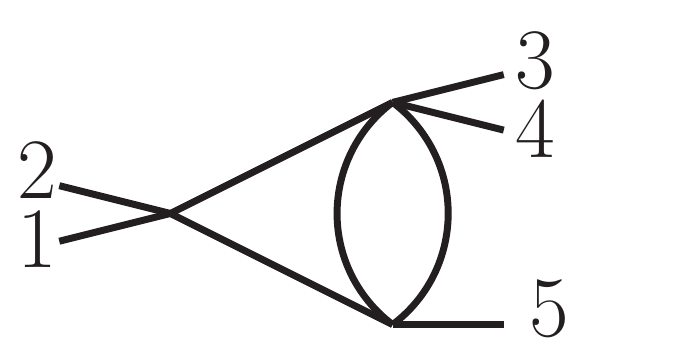} & 
  \raisebox{.2cm}{$\mathcal{N}_{64}$} \\
 \footnotesize{$F[1,0,0,1, 0,1,0,1, 0,0,0]$} 
  & &
 \footnotesize{$F[1,0,1,0, 0,1,0,1, 0,0,0]$} &  \\\hline
 \includegraphics[width=2.5cm]{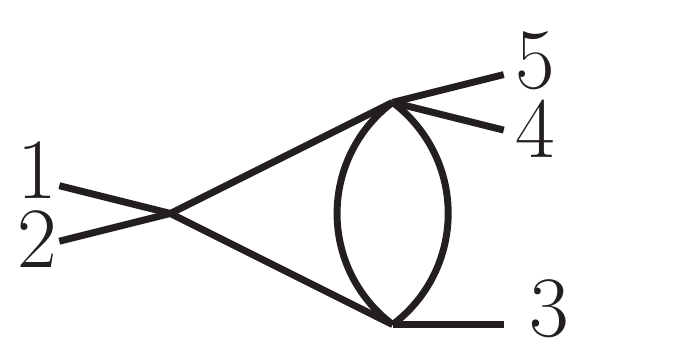} &
  \raisebox{.2cm}{$\mathcal{N}_{65}$} &
 \includegraphics[width=2.5cm]{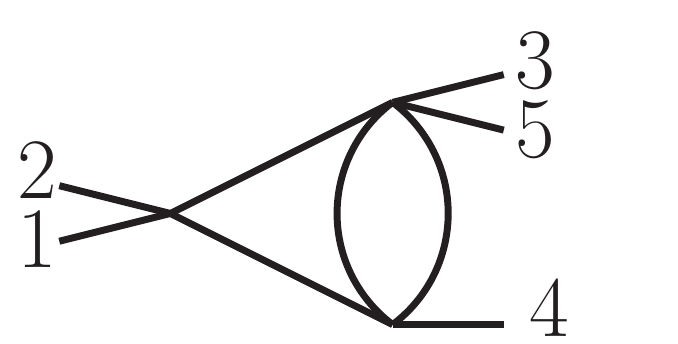} &
  \raisebox{.2cm}{$\mathcal{N}_{66}$}  \\
 \footnotesize{$F[1,0,1,0, 1,0,0,1, 0,0,0]$} 
  & &
 \footnotesize{$F[1,0,1,0, 1,0,1,0, 0,0,0]$} &  \\\hline
 \includegraphics[width=2cm]{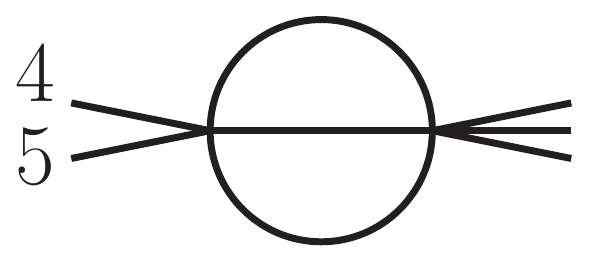} &
  \raisebox{0cm}{$\mathcal{N}_{67}$} &
 \includegraphics[width=2cm]{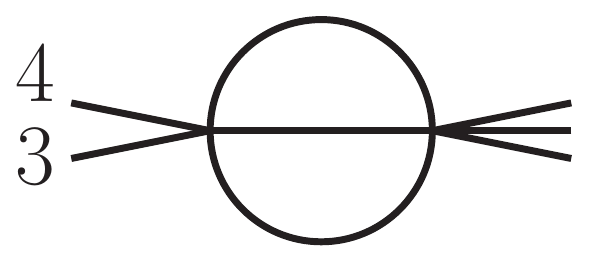} &
  \raisebox{0cm}{$\mathcal{N}_{68}$}  \\
 \footnotesize{$F[0,0,0,1, 0,1,1,0, 0,0,0]$} 
  & &
 \footnotesize{$F[0,0,1,0, 0,1,0,1, 0,0,0]$} &  \\\hline
 \includegraphics[width=2cm]{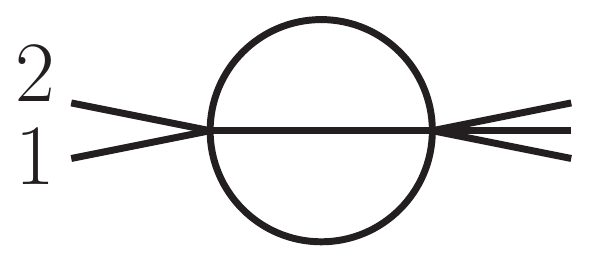}  &
  \raisebox{.0cm}{$\mathcal{N}_{69}$} &
 \includegraphics[width=2cm]{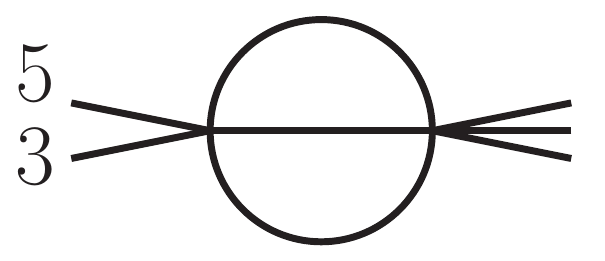} &
  \raisebox{.0cm}{$\mathcal{N}_{70}$} \\
 \footnotesize{$F[0,0,1,0, 0,1,1,0, 0,0,0]$} 
  & &
 \footnotesize{$F[0,0,1,0, 1,0,1,0, 0,0,0]$} &  \\\hline
 \includegraphics[width=2cm]{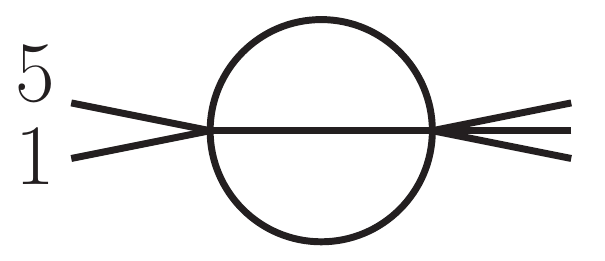} &
  \raisebox{.0cm}{$\mathcal{N}_{71}$} &
 \includegraphics[width=2cm]{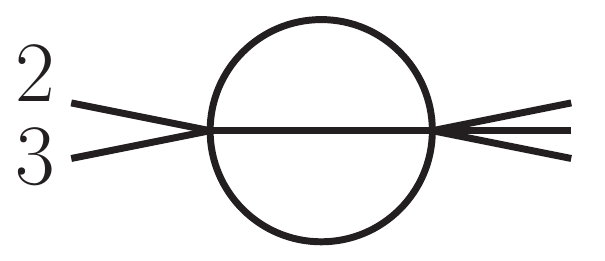} &
  \raisebox{.0cm}{$\mathcal{N}_{72}$}  \\ 
 \footnotesize{$F[0,1,0,0, 0,1,0,1, 0,0,0]$} 
  & &
 \footnotesize{$F[0,1,0,0, 1,0,0,1, 0,0,0]$} &  \\\hline
 \includegraphics[width=2cm]{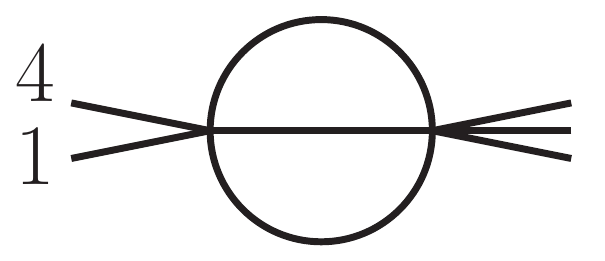} &
  \raisebox{.0cm}{$\mathcal{N}_{73}$} \\
 \footnotesize{$F[0,1,0,0, 1,0,1,0, 0,0,0]$} &
 
 \end{tabular}
 \caption{Diagrammatic representation of each sector and
 associated numerators.}
 \label{tab:first3}
\end{table}
\clearpage

\bibliographystyle{JHEP}
\bibliography{nonplanar}

\providecommand{\href}[2]{#2}\begingroup\raggedright\begin{thebibliography}{10}

\bibitem{Henn:2016jdu}
J.~M. Henn and B.~Mistlberger, \emph{{Four-Gluon Scattering at Three Loops,
  Infrared Structure, and the Regge Limit}},
  \href{http://dx.doi.org/10.1103/PhysRevLett.117.171601}{\emph{Phys. Rev.
  Lett.} {\bf 117} (2016) 171601},
  [\href{https://arxiv.org/abs/1608.00850}{{\tt 1608.00850}}].

\bibitem{Carrasco:2011mn}
J.~J. Carrasco and H.~Johansson, \emph{{Five-Point Amplitudes in N=4
  Super-Yang-Mills Theory and N=8 Supergravity}},
  \href{http://dx.doi.org/10.1103/PhysRevD.85.025006}{\emph{Phys. Rev.} {\bf
  D85} (2012) 025006}, [\href{https://arxiv.org/abs/1106.4711}{{\tt
  1106.4711}}].

\bibitem{Bern:2015ple}
Z.~Bern, E.~Herrmann, S.~Litsey, J.~Stankowicz and J.~Trnka, \emph{{Evidence
  for a Nonplanar Amplituhedron}},
  \href{http://dx.doi.org/10.1007/JHEP06(2016)098}{\emph{JHEP} {\bf 06} (2016)
  098}, [\href{https://arxiv.org/abs/1512.08591}{{\tt 1512.08591}}].

\bibitem{Badger:2013gxa}
S.~Badger, H.~Frellesvig and Y.~Zhang, \emph{{A Two-Loop Five-Gluon Helicity
  Amplitude in QCD}},
  \href{http://dx.doi.org/10.1007/JHEP12(2013)045}{\emph{JHEP} {\bf 12} (2013)
  045}, [\href{https://arxiv.org/abs/1310.1051}{{\tt 1310.1051}}].

\bibitem{Gehrmann:2015bfy}
T.~Gehrmann, J.~M. Henn and N.~A. Lo~Presti, \emph{{Analytic form of the
  two-loop planar five-gluon all-plus-helicity amplitude in QCD}},
  \href{http://dx.doi.org/10.1103/PhysRevLett.116.189903,
  10.1103/PhysRevLett.116.062001}{\emph{Phys. Rev. Lett.} {\bf 116} (2016)
  062001}, [\href{https://arxiv.org/abs/1511.05409}{{\tt 1511.05409}}].

\bibitem{Dunbar:2016aux}
D.~C. Dunbar and W.~B. Perkins, \emph{{Two-loop five-point all plus helicity
  Yang-Mills amplitude}},
  \href{http://dx.doi.org/10.1103/PhysRevD.93.085029}{\emph{Phys. Rev.} {\bf
  D93} (2016) 085029}, [\href{https://arxiv.org/abs/1603.07514}{{\tt
  1603.07514}}].

\bibitem{Badger:2017jhb}
S.~Badger, C.~Br{\o}nnum-Hansen, H.~B. Hartanto and T.~Peraro, \emph{{First
  look at two-loop five-gluon scattering in QCD}},
  \href{http://dx.doi.org/10.1103/PhysRevLett.120.092001}{\emph{Phys. Rev.
  Lett.} {\bf 120} (2018) 092001},
  [\href{https://arxiv.org/abs/1712.02229}{{\tt 1712.02229}}].

\bibitem{Abreu:2017hqn}
S.~Abreu, F.~Febres~Cordero, H.~Ita, B.~Page and M.~Zeng, \emph{{Planar
  Two-Loop Five-Gluon Amplitudes from Numerical Unitarity}},
  \href{http://dx.doi.org/10.1103/PhysRevD.97.116014}{\emph{Phys. Rev.} {\bf
  D97} (2018) 116014}, [\href{https://arxiv.org/abs/1712.03946}{{\tt
  1712.03946}}].

\bibitem{Bern:1994zx}
Z.~Bern, L.~J. Dixon, D.~C. Dunbar and D.~A. Kosower, \emph{{One loop n point
  gauge theory amplitudes, unitarity and collinear limits}},
  \href{http://dx.doi.org/10.1016/0550-3213(94)90179-1}{\emph{Nucl. Phys.} {\bf
  B425} (1994) 217--260}, [\href{https://arxiv.org/abs/hep-ph/9403226}{{\tt
  hep-ph/9403226}}].

\bibitem{Bern:1994cg}
Z.~Bern, L.~J. Dixon, D.~C. Dunbar and D.~A. Kosower, \emph{{Fusing gauge
  theory tree amplitudes into loop amplitudes}},
  \href{http://dx.doi.org/10.1016/0550-3213(94)00488-Z}{\emph{Nucl. Phys.} {\bf
  B435} (1995) 59--101}, [\href{https://arxiv.org/abs/hep-ph/9409265}{{\tt
  hep-ph/9409265}}].

\bibitem{Bern:1997sc}
Z.~Bern, L.~J. Dixon and D.~A. Kosower, \emph{{One loop amplitudes for e+ e- to
  four partons}},
  \href{http://dx.doi.org/10.1016/S0550-3213(97)00703-7}{\emph{Nucl. Phys.}
  {\bf B513} (1998) 3--86}, [\href{https://arxiv.org/abs/hep-ph/9708239}{{\tt
  hep-ph/9708239}}].

\bibitem{Britto:2004nc}
R.~Britto, F.~Cachazo and B.~Feng, \emph{{Generalized unitarity and one-loop
  amplitudes in N=4 super-Yang-Mills}},
  \href{http://dx.doi.org/10.1016/j.nuclphysb.2005.07.014}{\emph{Nucl. Phys.}
  {\bf B725} (2005) 275--305},
  [\href{https://arxiv.org/abs/hep-th/0412103}{{\tt hep-th/0412103}}].

\bibitem{Chetyrkin:1981qh}
K.~G. Chetyrkin and F.~V. Tkachov, \emph{{Integration by Parts: The Algorithm
  to Calculate beta Functions in 4 Loops}},
  \href{http://dx.doi.org/10.1016/0550-3213(81)90199-1}{\emph{Nucl. Phys.} {\bf
  B192} (1981) 159--204}.

\bibitem{Laporta:2001dd}
S.~Laporta, \emph{{High precision calculation of multiloop Feynman integrals by
  difference equations}},
  \href{http://dx.doi.org/10.1016/S0217-751X(00)00215-7,
  10.1142/S0217751X00002157}{\emph{Int. J. Mod. Phys.} {\bf A15} (2000)
  5087--5159}, [\href{https://arxiv.org/abs/hep-ph/0102033}{{\tt
  hep-ph/0102033}}].

\bibitem{Laporta:1996mq}
S.~Laporta and E.~Remiddi, \emph{{The Analytical value of the electron (g-2) at
  order alpha**3 in QED}},
  \href{http://dx.doi.org/10.1016/0370-2693(96)00439-X}{\emph{Phys. Lett.} {\bf
  B379} (1996) 283--291}, [\href{https://arxiv.org/abs/hep-ph/9602417}{{\tt
  hep-ph/9602417}}].

\bibitem{Papadopoulos:2015jft}
C.~G. Papadopoulos, D.~Tommasini and C.~Wever, \emph{{The Pentabox Master
  Integrals with the Simplified Differential Equations approach}},
  \href{http://dx.doi.org/10.1007/JHEP04(2016)078}{\emph{JHEP} {\bf 04} (2016)
  078}, [\href{https://arxiv.org/abs/1511.09404}{{\tt 1511.09404}}].

\bibitem{Gehrmann:2018yef}
T.~Gehrmann, J.~M. Henn and N.~A.~L. Presti, \emph{{Pentagon functions for
  massless planar scattering amplitudes}},
  \href{https://arxiv.org/abs/1807.09812}{{\tt 1807.09812}}.

\bibitem{Chicherin:2017dob}
D.~Chicherin, J.~Henn and V.~Mitev, \emph{{Bootstrapping pentagon functions}},
  \href{http://dx.doi.org/10.1007/JHEP05(2018)164}{\emph{JHEP} {\bf 05} (2018)
  164}, [\href{https://arxiv.org/abs/1712.09610}{{\tt 1712.09610}}].

\bibitem{Chicherin:2018ubl}
D.~Chicherin, J.~M. Henn and E.~Sokatchev, \emph{{Amplitudes from
  superconformal Ward identities}},
  \href{http://dx.doi.org/10.1103/PhysRevLett.121.021602}{\emph{Phys. Rev.
  Lett.} {\bf 121} (2018) 021602},
  [\href{https://arxiv.org/abs/1804.03571}{{\tt 1804.03571}}].

\bibitem{Bern:2018oao}
Z.~Bern, M.~Enciso, C.-H. Shen and M.~Zeng, \emph{{Dual Conformal Structure
  Beyond the Planar Limit}},  \href{https://arxiv.org/abs/1806.06509}{{\tt
  1806.06509}}.

\bibitem{Chicherin:2018wes}
D.~Chicherin, J.~M. Henn and E.~Sokatchev, \emph{{Implications of nonplanar
  dual conformal symmetry}},  \href{https://arxiv.org/abs/1807.06321}{{\tt
  1807.06321}}.

\bibitem{Kotikov:1990kg}
A.~V. Kotikov, \emph{{Differential equations method: New technique for massive
  Feynman diagrams calculation}},
  \href{http://dx.doi.org/10.1016/0370-2693(91)90413-K}{\emph{Phys. Lett.} {\bf
  B254} (1991) 158--164}.

\bibitem{Bern:1993kr}
Z.~Bern, L.~J. Dixon and D.~A. Kosower, \emph{{Dimensionally regulated pentagon
  integrals}},
  \href{http://dx.doi.org/10.1016/0550-3213(94)90398-0}{\emph{Nucl. Phys.} {\bf
  B412} (1994) 751--816}, [\href{https://arxiv.org/abs/hep-ph/9306240}{{\tt
  hep-ph/9306240}}].

\bibitem{Remiddi:1997ny}
E.~Remiddi, \emph{{Differential equations for Feynman graph amplitudes}},
  {\emph{Nuovo Cim.} {\bf A110} (1997) 1435--1452},
  [\href{https://arxiv.org/abs/hep-th/9711188}{{\tt hep-th/9711188}}].

\bibitem{Gehrmann:1999as}
T.~Gehrmann and E.~Remiddi, \emph{{Differential equations for two loop four
  point functions}},
  \href{http://dx.doi.org/10.1016/S0550-3213(00)00223-6}{\emph{Nucl. Phys.}
  {\bf B580} (2000) 485--518},
  [\href{https://arxiv.org/abs/hep-ph/9912329}{{\tt hep-ph/9912329}}].

\bibitem{Argeri:2007up}
M.~Argeri and P.~Mastrolia, \emph{{Feynman Diagrams and Differential
  Equations}}, \href{http://dx.doi.org/10.1142/S0217751X07037147}{\emph{Int. J.
  Mod. Phys.} {\bf A22} (2007) 4375--4436},
  [\href{https://arxiv.org/abs/0707.4037}{{\tt 0707.4037}}].

\bibitem{Henn:2013pwa}
J.~M. Henn, \emph{{Multiloop integrals in dimensional regularization made
  simple}}, \href{http://dx.doi.org/10.1103/PhysRevLett.110.251601}{\emph{Phys.
  Rev. Lett.} {\bf 110} (2013) 251601},
  [\href{https://arxiv.org/abs/1304.1806}{{\tt 1304.1806}}].

\bibitem{Boels:2018nrr}
R.~H. Boels, Q.~Jin and H.~Luo, \emph{{Efficient integrand reduction for
  particles with spin}},  \href{https://arxiv.org/abs/1802.06761}{{\tt
  1802.06761}}.

\bibitem{Chawdhry:2018awn}
H.~A. Chawdhry, M.~A. Lim and A.~Mitov, \emph{{Two-loop five-point massless QCD
  amplitudes within the IBP approach}},
  \href{https://arxiv.org/abs/1805.09182}{{\tt 1805.09182}}.

\bibitem{Anastasiou:2004vj}
C.~Anastasiou and A.~Lazopoulos, \emph{{Automatic integral reduction for higher
  order perturbative calculations}},
  \href{http://dx.doi.org/10.1088/1126-6708/2004/07/046}{\emph{JHEP} {\bf 07}
  (2004) 046}, [\href{https://arxiv.org/abs/hep-ph/0404258}{{\tt
  hep-ph/0404258}}].

\bibitem{vonManteuffel:2012np}
A.~von Manteuffel and C.~Studerus, \emph{{Reduze 2 - Distributed Feynman
  Integral Reduction}},  \href{https://arxiv.org/abs/1201.4330}{{\tt
  1201.4330}}.

\bibitem{Lee:2012cn}
R.~N. Lee, \emph{{Presenting LiteRed: a tool for the Loop InTEgrals
  REDuction}},  \href{https://arxiv.org/abs/1212.2685}{{\tt 1212.2685}}.

\bibitem{Smirnov:2014hma}
A.~V. Smirnov, \emph{{FIRE5: a C++ implementation of Feynman Integral
  REduction}}, \href{http://dx.doi.org/10.1016/j.cpc.2014.11.024}{\emph{Comput.
  Phys. Commun.} {\bf 189} (2015) 182--191},
  [\href{https://arxiv.org/abs/1408.2372}{{\tt 1408.2372}}].

\bibitem{Maierhoefer:2017hyi}
P.~Maierh{\"o}fer, J.~Usovitsch and P.~Uwer, \emph{{Kira---A Feynman integral
  reduction program}},
  \href{http://dx.doi.org/10.1016/j.cpc.2018.04.012}{\emph{Comput. Phys.
  Commun.} {\bf 230} (2018) 99--112},
  [\href{https://arxiv.org/abs/1705.05610}{{\tt 1705.05610}}].

\bibitem{Gluza:2010ws}
J.~Gluza, K.~Kajda and D.~A. Kosower, \emph{{Towards a Basis for Planar
  Two-Loop Integrals}},
  \href{http://dx.doi.org/10.1103/PhysRevD.83.045012}{\emph{Phys. Rev.} {\bf
  D83} (2011) 045012}, [\href{https://arxiv.org/abs/1009.0472}{{\tt
  1009.0472}}].

\bibitem{Ita:2015tya}
H.~Ita, \emph{{Two-loop Integrand Decomposition into Master Integrals and
  Surface Terms}},
  \href{http://dx.doi.org/10.1103/PhysRevD.94.116015}{\emph{Phys. Rev.} {\bf
  D94} (2016) 116015}, [\href{https://arxiv.org/abs/1510.05626}{{\tt
  1510.05626}}].

\bibitem{Larsen:2015ped}
K.~J. Larsen and Y.~Zhang, \emph{{Integration-by-parts reductions from
  unitarity cuts and algebraic geometry}},
  \href{http://dx.doi.org/10.1103/PhysRevD.93.041701}{\emph{Phys. Rev.} {\bf
  D93} (2016) 041701}, [\href{https://arxiv.org/abs/1511.01071}{{\tt
  1511.01071}}].

\bibitem{Georgoudis:2016wff}
A.~Georgoudis, K.~J. Larsen and Y.~Zhang, \emph{{Azurite: An algebraic geometry
  based package for finding bases of loop integrals}},
  \href{http://dx.doi.org/10.1016/j.cpc.2017.08.013}{\emph{Comput. Phys.
  Commun.} {\bf 221} (2017) 203--215},
  [\href{https://arxiv.org/abs/1612.04252}{{\tt 1612.04252}}].

\bibitem{Abreu:2017xsl}
S.~Abreu, F.~Febres~Cordero, H.~Ita, M.~Jaquier, B.~Page and M.~Zeng,
  \emph{{Two-Loop Four-Gluon Amplitudes from Numerical Unitarity}},
  \href{http://dx.doi.org/10.1103/PhysRevLett.119.142001}{\emph{Phys. Rev.
  Lett.} {\bf 119} (2017) 142001},
  [\href{https://arxiv.org/abs/1703.05273}{{\tt 1703.05273}}].

\bibitem{Boehm:2017wjc}
J.~Boehm, A.~Georgoudis, K.~J. Larsen, M.~Schulze and Y.~Zhang, \emph{{Complete
  sets of logarithmic vector fields for integration-by-parts identities of
  Feynman integrals}},  \href{https://arxiv.org/abs/1712.09737}{{\tt
  1712.09737}}.

\bibitem{Boehm:2018fpv}
J.~Boehm, A.~Georgoudis, K.~J. Larsen, H.~Schönemann and Y.~Zhang,
  \emph{{Complete integration-by-parts reductions of the non-planar hexagon-box
  via module intersections}},  \href{https://arxiv.org/abs/1805.01873}{{\tt
  1805.01873}}.

\bibitem{Frellesvig:2017aai}
H.~Frellesvig and C.~G. Papadopoulos, \emph{{Cuts of Feynman Integrals in
  Baikov representation}},
  \href{http://dx.doi.org/10.1007/JHEP04(2017)083}{\emph{JHEP} {\bf 04} (2017)
  083}, [\href{https://arxiv.org/abs/1701.07356}{{\tt 1701.07356}}].

\bibitem{Zeng:2017ipr}
M.~Zeng, \emph{{Differential equations on unitarity cut surfaces}},
  \href{http://dx.doi.org/10.1007/JHEP06(2017)121}{\emph{JHEP} {\bf 06} (2017)
  121}, [\href{https://arxiv.org/abs/1702.02355}{{\tt 1702.02355}}].

\bibitem{Bosma:2017hrk}
J.~Bosma, K.~J. Larsen and Y.~Zhang, \emph{{Differential equations for loop
  integrals in Baikov representation}},
  \href{http://dx.doi.org/10.1103/PhysRevD.97.105014}{\emph{Phys. Rev.} {\bf
  D97} (2018) 105014}, [\href{https://arxiv.org/abs/1712.03760}{{\tt
  1712.03760}}].

\bibitem{Goncharov:2010jf}
A.~B. Goncharov, M.~Spradlin, C.~Vergu and A.~Volovich, \emph{{Classical
  Polylogarithms for Amplitudes and Wilson Loops}},
  \href{http://dx.doi.org/10.1103/PhysRevLett.105.151605}{\emph{Phys. Rev.
  Lett.} {\bf 105} (2010) 151605}, [\href{https://arxiv.org/abs/1006.5703}{{\tt
  1006.5703}}].

\bibitem{Gaiotto:2011dt}
D.~Gaiotto, J.~Maldacena, A.~Sever and P.~Vieira, \emph{{Pulling the straps of
  polygons}}, \href{http://dx.doi.org/10.1007/JHEP12(2011)011}{\emph{JHEP} {\bf
  12} (2011) 011}, [\href{https://arxiv.org/abs/1102.0062}{{\tt 1102.0062}}].

\bibitem{Cutkosky:1960sp}
R.~E. Cutkosky, \emph{{Singularities and discontinuities of Feynman
  amplitudes}}, \href{http://dx.doi.org/10.1063/1.1703676}{\emph{J. Math.
  Phys.} {\bf 1} (1960) 429--433}.

\bibitem{Baikov:1996rk}
P.~A. Baikov, \emph{{Explicit solutions of the three loop vacuum integral
  recurrence relations}},
  \href{http://dx.doi.org/10.1016/0370-2693(96)00835-0}{\emph{Phys. Lett.} {\bf
  B385} (1996) 404--410}, [\href{https://arxiv.org/abs/hep-ph/9603267}{{\tt
  hep-ph/9603267}}].

\bibitem{Baikov:1996iu}
P.~A. Baikov, \emph{{Explicit solutions of the multiloop integral recurrence
  relations and its application}},
  \href{http://dx.doi.org/10.1016/S0168-9002(97)00126-5}{\emph{Nucl. Instrum.
  Meth.} {\bf A389} (1997) 347--349},
  [\href{https://arxiv.org/abs/hep-ph/9611449}{{\tt hep-ph/9611449}}].

\bibitem{Grozin:2011mt}
A.~G. Grozin, \emph{{Integration by parts: An Introduction}},
  \href{http://dx.doi.org/10.1142/S0217751X11053687}{\emph{Int. J. Mod. Phys.}
  {\bf A26} (2011) 2807--2854}, [\href{https://arxiv.org/abs/1104.3993}{{\tt
  1104.3993}}].

\bibitem{Zhang:2016kfo}
Y.~Zhang, \emph{{Lecture Notes on Multi-loop Integral Reduction and Applied
  Algebraic Geometry}},  2016.
\newblock \href{https://arxiv.org/abs/1612.02249}{{\tt 1612.02249}}.

\bibitem{Schabinger:2011dz}
R.~M. Schabinger, \emph{{A New Algorithm For The Generation Of
  Unitarity-Compatible Integration By Parts Relations}},
  \href{http://dx.doi.org/10.1007/JHEP01(2012)077}{\emph{JHEP} {\bf 01} (2012)
  077}, [\href{https://arxiv.org/abs/1111.4220}{{\tt 1111.4220}}].

\bibitem{Bern:2017gdk}
Z.~Bern, M.~Enciso, H.~Ita and M.~Zeng, \emph{{Dual Conformal Symmetry,
  Integration-by-Parts Reduction, Differential Equations and the Nonplanar
  Sector}}, \href{http://dx.doi.org/10.1103/PhysRevD.96.096017}{\emph{Phys.
  Rev.} {\bf D96} (2017) 096017}, [\href{https://arxiv.org/abs/1709.06055}{{\tt
  1709.06055}}].

\bibitem{DGPS}
W.~Decker, G.-M. Greuel, G.~Pfister and H.~Sch\"onemann, ``{\sc Singular}
  {4-1-1} --- {A} computer algebra system for polynomial computations.''
  \url{http://www.singular.uni-kl.de}, 2018.

\bibitem{lewis2008computer}
R.~H. Lewis, ``Computer algebra system fermat.''
  \url{http://home.bway.net/lewis/}, 2008.

\bibitem{Goncharov:1998}
A.~B. Goncharov, \emph{{Multiple polylogarithms, cyclotomy and modular
  complexes}}, {\emph{Math. Research Letters} {\bf 5(4)} (1998) 497}.

\bibitem{Goncharov:2001}
A.~B. Goncharov, \emph{{Multiple polylogarithms and mixed Tate motives}},
  {\emph{arXiv:math/0103059 [math.AG]} (2001) }.

\bibitem{Henn:2014qga}
J.~M. Henn, \emph{{Lectures on differential equations for Feynman integrals}},
  \href{http://dx.doi.org/10.1088/1751-8113/48/15/153001}{\emph{J. Phys.} {\bf
  A48} (2015) 153001}, [\href{https://arxiv.org/abs/1412.2296}{{\tt
  1412.2296}}].

\bibitem{Duhr:2014woa}
C.~Duhr, \emph{{Mathematical aspects of scattering amplitudes}},  in
  \emph{{Proceedings, Theoretical Advanced Study Institute in Elementary
  Particle Physics: Journeys Through the Precision Frontier: Amplitudes for
  Colliders (TASI 2014): Boulder, Colorado, June 2-27, 2014}}, pp.~419--476,
  2015.
\newblock \href{https://arxiv.org/abs/1411.7538}{{\tt 1411.7538}}.
\newblock \href{http://dx.doi.org/10.1142/9789814678766_0010}{DOI}.

\bibitem{Lee:2014ioa}
R.~N. Lee, \emph{{Reducing differential equations for multiloop master
  integrals}}, \href{http://dx.doi.org/10.1007/JHEP04(2015)108}{\emph{JHEP}
  {\bf 04} (2015) 108}, [\href{https://arxiv.org/abs/1411.0911}{{\tt
  1411.0911}}].

\bibitem{Gituliar:2017vzm}
O.~Gituliar and V.~Magerya, \emph{{Fuchsia: a tool for reducing differential
  equations for Feynman master integrals to epsilon form}},
  \href{http://dx.doi.org/10.1016/j.cpc.2017.05.004}{\emph{Comput. Phys.
  Commun.} {\bf 219} (2017) 329--338},
  [\href{https://arxiv.org/abs/1701.04269}{{\tt 1701.04269}}].

\bibitem{Prausa:2017ltv}
M.~Prausa, \emph{{epsilon: A tool to find a canonical basis of master
  integrals}}, \href{http://dx.doi.org/10.1016/j.cpc.2017.05.026}{\emph{Comput.
  Phys. Commun.} {\bf 219} (2017) 361--376},
  [\href{https://arxiv.org/abs/1701.00725}{{\tt 1701.00725}}].

\bibitem{Meyer:2017joq}
C.~Meyer, \emph{{Algorithmic transformation of multi-loop master integrals to a
  canonical basis with CANONICA}},
  \href{http://dx.doi.org/10.1016/j.cpc.2017.09.014}{\emph{Comput. Phys.
  Commun.} {\bf 222} (2018) 295--312},
  [\href{https://arxiv.org/abs/1705.06252}{{\tt 1705.06252}}].

\bibitem{Gehrmann:2001ck}
T.~Gehrmann and E.~Remiddi, \emph{{Two loop master integrals for gamma* $\to$ 3
  jets: The Nonplanar topologies}},
  \href{http://dx.doi.org/10.1016/S0550-3213(01)00074-8}{\emph{Nucl. Phys.}
  {\bf B601} (2001) 287--317},
  [\href{https://arxiv.org/abs/hep-ph/0101124}{{\tt hep-ph/0101124}}].

\bibitem{Kozlov:2015kol}
M.~G. Kozlov and R.~N. Lee, \emph{{One-loop pentagon integral in $d$ dimensions
  from differential equations in $\epsilon$-form}},
  \href{http://dx.doi.org/10.1007/JHEP02(2016)021}{\emph{JHEP} {\bf 02} (2016)
  021}, [\href{https://arxiv.org/abs/1512.01165}{{\tt 1512.01165}}].

\bibitem{Abreu:2018sat}
S.~Abreu, R.~Britto, C.~Duhr and E.~Gardi, \emph{{The diagrammatic coaction and
  the algebraic structure of cut Feynman integrals}},
  \href{http://dx.doi.org/10.22323/1.290.0002}{\emph{PoS} {\bf RADCOR2017}
  (2018) 002}, [\href{https://arxiv.org/abs/1803.05894}{{\tt 1803.05894}}].

\bibitem{Lee:2009dh}
R.~N. Lee, \emph{{Space-time dimensionality D as complex variable: Calculating
  loop integrals using dimensional recurrence relation and analytical
  properties with respect to D}},
  \href{http://dx.doi.org/10.1016/j.nuclphysb.2009.12.025}{\emph{Nucl.~Phys.}
  {\bf B830} (2010) 474--492}, [\href{https://arxiv.org/abs/0911.0252}{{\tt
  0911.0252}}].

\bibitem{Hodges:2009hk}
A.~Hodges, \emph{{Eliminating spurious poles from gauge-theoretic amplitudes}},
  \href{http://dx.doi.org/10.1007/JHEP05(2013)135}{\emph{JHEP} {\bf 05} (2013)
  135}, [\href{https://arxiv.org/abs/0905.1473}{{\tt 0905.1473}}].

\bibitem{Dixon:1996wi}
L.~J. Dixon, \emph{{Calculating scattering amplitudes efficiently}},  in
  \emph{{QCD and beyond. Proceedings, Theoretical Advanced Study Institute in
  Elementary Particle Physics, TASI-95, Boulder, USA, June 4-30, 1995}},
  pp.~539--584, 1996.
\newblock \href{https://arxiv.org/abs/hep-ph/9601359}{{\tt hep-ph/9601359}}.

\end{thebibliography}\endgroup
\end{document}